\documentclass{article}

\usepackage[dvipdfmx]{graphicx}
\usepackage{amsmath}        
\usepackage{amssymb}        
\usepackage{cite}
\usepackage[dvips]{color}
\usepackage{fancyhdr}
\usepackage[affil-it,blocks]{authblk}

\usepackage{ulem}
\usepackage{xcolor}

\makeatletter
 \renewcommand{\theequation}{%
 \thesection.\arabic{equation}}
 \@addtoreset{equation}{section}
\makeatother

\def\T{{\rm T}}

\def\GL{{G\kern-.12em L\kern-.04em}}
\def\OSp{{O\kern-.11em S\kern-.04em p}}
\def\IOSp{{I\kern-.06em O\kern-.11em S\kern-.04em p}}
\def\MN{{M\kern-.14em N}}
\def\NM{{N\kern-.14em M}}
\def\NL{{N\kern-.14em L}}
\def\LN{{L\kern-.11em N}}
\def\ML{{M\kern-.14em L}}
\def\LM{{L\kern-.11em M}}
\def\RN{{R\kern-.11em N}}
\def\NR{{N\kern-.14em R}}
\def\RM{{R\kern-.11em M}}
\def\MR{{M\kern-.14em R}}
\def\RL{{R\kern-.11em L}}
\def\LR{{L\kern-.11em R}}
\def\RS{{R\kern-.11em S}}
\def\SR{{S\kern-.11em R}}
\def\SN{{S\kern-.11em N}}
\def\NS{{N\kern-.11em S}}
\def\SM{{S\kern-.11em M}}
\def\MS{{M\kern-.11em S}}
\def\SL{{S\kern-.11em L}}
\def\LS{{L\kern-.11em S}}
\def\sqr#1#2{{\vcenter{\hrule height.#2pt
      \hbox{\vrule width.#2pt height#1pt \kern#1pt
          \vrule width.#2pt}
      \hrule height.#2pt}}}
\def\bra0{\langle0|}
\def\ket0{|0\rangle}
\def\soeji#1_#2#3{#1_{#2}\cdots#1_{#3}}
\def\longgLRarrow{\longleftarrow\kern-3pt\relbar\kern-3pt\relbar\kern-3pt%
\longrightarrow}
\def\longLRarrow{\longleftarrow\kern-3pt\relbar\kern-3pt\longrightarrow}
\def\longLarrow{\longleftarrow\kern-3pt\relbar\kern-3pt\relbar\kern-3pt\relbar}
\def\longRarrow{\relbar\kern-3pt\relbar\kern-3pt\relbar\kern-3pt\longrightarrow}
\def\bothDer#1#2#3{%
\overset{\kern-.7em\stackrel{#1}{#2}}{\partial_{#3}}}
 
\makeatletter
 \renewcommand{\theequation}{%
 \thesection.\arabic{equation}}
 \@addtoreset{equation}{section}
\makeatother

\begin{document}
\thispagestyle{fancy}

\title{BRST Formalism of  Weyl Conformal Gravity}


\author[1]{Ichiro Oda\thanks{ioda@sci.u-ryukyu.ac.jp}}
\author[2]{Philipp Saake\thanks{saake@mpi-hd.mpg.de}}
\affil[1]{Department of Physics, Faculty of Science, University of the Ryukyus, Nishihara, Okinawa 903-0213, Japan}
\affil[2]{Max-Planck-Institut f\"ur Kernphysik (MPIK), Saupfercheckweg 1, 69117 Heidelberg, Germany}

\date{}

\maketitle

\thispagestyle{fancy}

\begin{abstract}

We present the BRST formalism of a Weyl conformal gravity in Weyl geometry. Choosing the extended 
de Donder gauge-fixing condition (or harmonic gauge condition) for the general coordinate 
invariance and the new scalar gauge-fixing for the Weyl invariance we find that there is a Poincar${\rm{\acute{e}}}$-like 
$\IOSp(10|10)$ supersymmetry as in a Weyl invariant scalar-tensor gravity in Riemann
geometry. We also point out that there is a gravitational conformal symmetry in quantum gravity although there is
a massive Weyl gauge field as a result of spontaneous symmetry breakdown of Weyl gauge symmetry 
and account for how the gravitational conformal symmetry is spontaneously broken to the Poincar\'e symmetry. 
The corresponding massless Nambu-Goldstone bosons are the graviton and the dilaton. We also prove 
the unitarity of the physical S-matrix on the basis of the BRST quartet mechanism.

\end{abstract}

\section{Introduction}

More than one hundred years ago, H. Weyl has advocated a new theory to unify all the interactions
known at that time, namely gravitational interaction and electro-magnetic force, within the framework
of a newly established geometry which is nowadays called ``Weyl geometry'' \cite{Weyl, Scholz}.   
In Riemann geometry both length and angle are preserved under parallel transport while in Weyl geometry,
only angle, but not length, is preserved by the Weyl gauge field. Soon after the advent of the Weyl's idea, 
A. Einstein has criticized that regarding the spacing of atomic spectral lines, the prediction obtained 
from Weyl's theory and the experimental observations were in contradiction (this problem is sometimes called 
the second clock problem \cite{Penrose}), thus Weyl theory has been buried in oblivion 
for a long time.\footnote{Even during this period, there were some papers dealing with 
Weyl theory \cite{Dirac}-\!\cite{Cesare}.}

However, in recent years a considerate interest has been developed for Weyl conformal geometry. 
This is because it was found that the Weyl gauge field acquires a huge mass around the Planck scale and 
decouples at low energies, thereby avoiding the second clock problem \cite{Ghilencea1, Ghilencea2, Oda-P, Oda-Corf}. 
In addition, we have noticed the importance of global scale invariance and also local scale invariance, which is also 
called Weyl invariance, in formulating a theory beyond the Standard Model (BSM) \cite{Bardeen}
and quantum gravity. Hence, Weyl geometry provides us with a natural playground for describing Weyl symmetry.

The study of Weyl conformal gravity in Weyl geometry has been mainly limited to a classical analysis 
thus far.\footnote{At one-loop level, the effective potential has been already calculated in \cite{Oda-P, Oda-Corf}.}
One of motivations behind the present article is to present a quantum theory of Weyl conformal gravity.   
To this end, we construct a BRST formalism of the theory from which we can shed some light on important features 
of quantum aspects of Weyl conformal gravity in Weyl geometry. For instance, as been already shown in case of
Weyl invariant scalar-tensor gravity \cite{Oda-Q, Oda-W, Oda-V}, there is an extended $\IOSp(10|10)$ choral 
symmetry compared with the $\IOSp(8|8)$ choral symmetry in Einstein's gravity \cite{Nakanishi, N-O-text}.
This extended symmetry is not confined to the sector of the Nakanishi-Lautrup auxiliary fields 
and the Faddeev-Popov (FP) (anti-)ghosts but relevant to a classical theory. Moreover, it can be shown 
that we have a gravitational analog of conformal algebra as a subalgebra of the $\IOSp(10|10)$ choral symmetry. 
That algebra then gives rise to a spontaneous symmetry breakdown to the Poincar\'e symmetry, 
by which we can prove that the graviton \cite{NO} and the dilaton  \cite{Oda-W, Oda-V} are exactly massless 
since they are the Nambu-Goldstone particles.

The paper is organized as follows. In Section 2, we give a brief review of Weyl geometry. In Section 3, we
consider a classical theory which is not only invariant under Weyl gauge transformation but also is
free of ghosts, that generally exist in the higher-derivative gravities.  Based on the classical theory in Section 3,
we fix the gauge symmetries by the extended de Donder gauge and new scalar gauge conditions and
construct a BRST invariant quantum Lagrangian in Section 4. In Section 5, we perform the canonical quantization
of the quantum Lagrangian where we meet primary and secondary constraints associated with Weyl 
symmetry. They are the second-class constraints and hence are treated by applying the Dirac brackets.
In Section 6, we prove the unitarity of the physical S-matrix on the basis of the BRST quartet mechanism.
We find that physical modes are the two polarizations of the massless graviton and the three modes
of the massive Weyl gauge fields. Furthermore, it is shown that the massless dilaton, which is eaten by the
Weyl gauge field via the Higgs mechanism, belongs to the unphysical sector. In Section 7, we show that
the quantum Lagrangian of Weyl conformal gravity possesses the huge global $\IOSp(10|10)$ choral symmetry. 
In Section 8, we point out that there exists a gravitational analog of conformal
symmetry in quantum gravity and investigate the spontaneous symmetry breaking. We find that the
graviton and the dilaton are massless Nambu-Goldstone bosons.
The final section is devoted to discussion.

\section{Review of Weyl conformal geometry} 

In this section, we briefly review the basic concepts and definitions of Weyl conformal 
geometry \cite{Oda-Corf}.\footnote{We follow 
the notation and conventions of MTW textbook \cite{MTW}. Lower case Greek letters $\mu, \nu, \cdots$ and 
Latin ones $i, j, \cdots$ are used for space-time and spatial indices, respectively; for instance, 
$\mu= 0, 1, 2, 3$ and $i = 1, 2, 3$. The Riemann curvature tensor and the Ricci tensor are respectively 
defined by $R^\rho{}_{\sigma\mu\nu} = \partial_\mu \Gamma^\rho_{\sigma\nu} 
- \partial_\nu \Gamma^\rho_{\sigma\mu} + \Gamma^\rho_{\lambda\mu} \Gamma^\lambda_{\sigma\nu} 
- \Gamma^\rho_{\lambda\nu} \Gamma^\lambda_{\sigma\mu}$ and $R_{\mu\nu} = R^\rho{}_{\mu\rho\nu}$. 
The Minkowski metric tensor is denoted by $\eta_{\mu\nu}$; $\eta_{00} = - \eta_{11} = - \eta_{22} 
= - \eta_{33} = -1$ and $\eta_{\mu\nu} = 0$ for $\mu \neq \nu$.} 
In Weyl geometry, the Weyl gauge transformation, which is the sum of a local scale transformation for 
a generic field $\Phi (x)$ and a gauge transformation for the Weyl gauge field $S_\mu(x)$, is defined as
\begin{eqnarray}
\Phi (x) \rightarrow \Phi^\prime (x) = e^{w \Lambda(x)} \Phi (x), \qquad
S_\mu (x) \rightarrow S^\prime_\mu (x) = S_\mu (x) - \frac{1}{f} \partial_\mu \Lambda (x),
\label{Weyl transf}
\end{eqnarray}
where $w$ is called the ``Weyl weight'', or simply ``weight'' henceforth, $f$ is the coupling constant 
for the non-compact Abelian gauge group, and $\Lambda(x)$ is a local parameter for the Weyl transformation. 
The Weyl gauge transformation for various fields is explicitly given by
\begin{eqnarray}
g_{\mu\nu} (x) &\rightarrow& g_{\mu\nu}^\prime (x) = e^{2 \Lambda(x)} g_{\mu\nu}(x), \qquad
\phi (x) \rightarrow \phi^\prime (x) = e^{- \Lambda(x)} \phi (x),  \nonumber\\
\psi (x) &\rightarrow& \psi^\prime (x) = e^{- \frac{3}{2} \Lambda(x)} \psi (x), \qquad
A_\mu (x) \rightarrow A^\prime_\mu (x) = A_\mu (x),
\label{Weyl transf 2}
\end{eqnarray}
where $g_{\mu\nu} (x)$, $\phi (x)$, $\psi (x)$ and $A_\mu (x)$ are the metric tensor, scalar, spinor,
and electromagnetic gauge fields, respectively. The covariant derivative $D_\mu$ for the Weyl gauge
transformation for a generic field $\Phi (x)$ of weight $w$ is defined as
\begin{eqnarray}
D_\mu \Phi \equiv \partial_\mu \Phi + w f S_\mu \Phi,
\label{W-cov-deriv}
\end{eqnarray}
which transforms covariantly under the Weyl transformation:
\begin{eqnarray}
D_\mu \Phi \rightarrow (D_\mu \Phi)^\prime = e^{w \Lambda(x)} D_\mu \Phi.
\label{S-cov-transf}
\end{eqnarray}

The Weyl geometry is defined as a geometry with a real symmetric metric tensor $g_{\mu\nu}
(= g_{\nu\mu})$ and a symmetric connection $\tilde \Gamma^\lambda_{\mu\nu} (= \tilde \Gamma^\lambda_{\nu\mu})$ 
which is defined as\footnote{We often use the tilde characters to express quantities belonging to Weyl geometry.}
\begin{eqnarray}
\tilde \Gamma^\lambda_{\mu\nu} &=& \frac{1}{2} g^{\lambda\rho} \left( D_\mu g_{\nu\rho} + D_\nu g_{\mu\rho}
- D_\rho g_{\mu\nu} \right)
\nonumber\\
&=& \Gamma^\lambda_{\mu\nu} + f \left( S_\mu \delta^\lambda_\nu + S_\nu \delta^\lambda_\mu 
- S^\lambda g_{\mu\nu} \right),
\label{W-connection}
\end{eqnarray}
where $\Gamma^\lambda_{\mu\nu}$ is the standard Christoffel symbol in Riemann geometry. The most important 
difference between Riemann geometry and Weyl geometry lies in the fact that in Riemann geometry 
the metric condition is satisfied:
\begin{eqnarray}
\nabla_\lambda g_{\mu\nu} \equiv \partial_\lambda g_{\mu\nu} - \Gamma^\rho_{\lambda\mu} 
g_{\rho\nu} - \Gamma^\rho_{\lambda\nu} g_{\mu\rho} = 0, 
\label{Metric cond}
\end{eqnarray}
while in Weyl geometry we have:
\begin{eqnarray}
\tilde \nabla_\lambda g_{\mu\nu} \equiv \partial_\lambda g_{\mu\nu} - \tilde \Gamma^\rho_{\lambda\mu} 
g_{\rho\nu} - \tilde \Gamma^\rho_{\lambda\nu} g_{\mu\rho}
= - 2 f S_\lambda g_{\mu\nu},
\label{W-metric cond}
\end{eqnarray}
where $\nabla_\mu$ and $\tilde \nabla_\mu$ are covariant derivatives for diffeomorphisms in Riemann 
and Weyl geometries, respectively. Since the metric condition (\ref{Metric cond}) implies that both length 
and angle are preserved under parallel transport, Eq. (\ref{W-metric cond}) shows that only angle, 
but not length, is preserved by the Weyl connection.

The general covariant derivative for both diffeomorphisms and Weyl gauge transformation, for instance,
for a covariant vector of weight $w$, is defined as
\begin{eqnarray}
{\cal D}_\mu V_\nu &\equiv& D_\mu V_\nu - \tilde \Gamma^\rho_{\mu\nu} V_\rho  \nonumber\\
&=& \tilde \nabla_\mu V_\nu + w f S_\mu V_\nu \nonumber\\
&=& \nabla_\mu V_\nu + w f S_\mu V_\nu - f ( S_\mu \delta^\rho _\nu + S_\nu \delta^\rho _\mu
- S^\rho g_{\mu\nu} ) V_\rho \nonumber\\
&=& \partial_\mu V_\nu + w f S_\mu V_\nu - \Gamma^\rho_{\mu\nu} V_\rho
- f ( S_\mu \delta^\rho _\nu + S_\nu \delta^\rho _\mu - S^\rho g_{\mu\nu} ) V_\rho.
\label{Gen-cov-deriv}
\end{eqnarray}
One can verify that using the general covariant derivative, the following metric condition is 
satisfied:
\begin{eqnarray}
{\cal D}_\lambda g_{\mu\nu} = 0.
\label{Gen-metric cond}
\end{eqnarray}
Moreover, under Weyl gauge transformation the general covariant derivative for a generic field 
$\Phi$ of weight $w$ transforms in a covariant manner as desired:
\begin{eqnarray}
{\cal D}_\mu \Phi \rightarrow ({\cal D}_\mu \Phi)^\prime = e^{w \Lambda(x)} {\cal D}_\mu \Phi,
\label{Gen-cov-transf}
\end{eqnarray}
because the Weyl connection is invariant under Weyl gauge transformation, i.e., 
$\tilde \Gamma^{\prime \rho}_{\mu\nu} = \tilde \Gamma^\rho_{\mu\nu}$.

As in Riemann geometry, in Weyl geometry one can also construct a Weyl invariant curvature
tensor $\tilde R_{\mu\nu\rho} \, ^\sigma$ via a commutator of the covariant 
derivative $\tilde \nabla_\mu$:
\begin{eqnarray}
[ \tilde \nabla_\mu, \tilde \nabla_\nu ] V_\rho = \tilde R_{\mu\nu\rho} \, ^\sigma V_\sigma.
\label{Commutator}
\end{eqnarray}
Calculating this commutator, one finds that
\begin{eqnarray}
\tilde R_{\mu\nu\rho} \, ^\sigma &=& \partial_\nu \tilde \Gamma^\sigma_{\mu\rho} 
- \partial_\mu \tilde \Gamma^\sigma_{\nu\rho} + \tilde \Gamma^\alpha_{\mu\rho} \tilde \Gamma^\sigma_{\alpha\nu} 
- \tilde \Gamma^\alpha_{\nu\rho} \tilde \Gamma^\sigma_{\alpha\mu}
\nonumber\\
&=& R_{\mu\nu\rho} \, ^\sigma + 2 f \left( \delta^\sigma_{[\mu} \nabla_{\nu]} S_\rho 
- \delta^\sigma_\rho \nabla_{[\mu} S_{\nu]} - g_{\rho [\mu} \nabla_{\nu]} S^\sigma \right)
\nonumber\\
&+& 2 f^2 \left( S_{[\mu} \delta^\sigma_{\nu]} S_\rho - S_{[\mu} g_{\nu]\rho} S^\sigma
+ \delta^\sigma_{[\mu} g_{\nu]\rho} S_\alpha S^\alpha \right),
\label{W-curv-tensor}
\end{eqnarray}
where $R_{\mu\nu\rho} \, ^\sigma$ is the curvature tensor in Riemann geometry and
we have defined the antisymmetrization by the square bracket, i.e., $A_{[\mu} B_{\nu]} \equiv 
\frac{1}{2} ( A_\mu B_\nu - A_\nu B_\mu )$. Then, it is straightforward to prove the following identities:
\begin{eqnarray}
\tilde R_{\mu\nu\rho} \, ^\sigma = - \tilde R_{\nu\mu\rho} \, ^\sigma,  \qquad
\tilde R_{[\mu\nu\rho]} \, ^\sigma = 0, \qquad
\tilde \nabla_{[\lambda} \tilde R_{\mu\nu]\rho} \, ^\sigma = 0.
\label{W-curv-identity}
\end{eqnarray}

From $\tilde R_{\mu\nu\rho} \, ^\sigma$ one can define a Weyl invariant 
Ricci tensor:
\begin{eqnarray}
\tilde R_{\mu\nu} &\equiv& \tilde R_{\mu\rho\nu} \, ^\rho
\nonumber\\
&=& R_{\mu\nu} + f \left( - 2 \nabla_\mu S_\nu - H_{\mu\nu} - g_{\mu\nu} \nabla_{\alpha} S^\alpha \right)
\nonumber\\
&+& 2 f^2 \left( S_\mu S_\nu - g_{\mu\nu} S_\alpha S^\alpha \right).
\label{W-Ricci-tensor}
\end{eqnarray}
Let us note that 
\begin{eqnarray}
\tilde R_{[\mu\nu]} \equiv \frac{1}{2} ( \tilde R_{\mu\nu} - \tilde R_{\nu\mu} ) = - 2 f H_{\mu\nu}.
\label{W-Ricci-tensor 2}
\end{eqnarray}
Similarly, one can define not a Weyl invariant but a Weyl covariant scalar curvature:
\begin{eqnarray}
\tilde R \equiv g^{\mu\nu} \tilde R_{\mu\nu} 
= R - 6 f \nabla_\mu S^\mu - 6 f^2 S_\mu S^\mu.
\label{W-scalar-curv}
\end{eqnarray}
One finds that under Weyl gauge transformation, $\tilde R \rightarrow \tilde R^\prime = e^{- 2 \Lambda(x)}
\tilde R$ while $\tilde \Gamma^\lambda_{\mu\nu}, \tilde R_{\mu\nu\rho} \, ^\sigma$ and $\tilde R_{\mu\nu}$
are all invariant.

We close this section by discussing a spinor field as an example of matter fields in Weyl geometry.  
As is well known, to describe a spinor field it is necessary to introduce the vierbein $e^a _\mu$, which is defined as
\begin{eqnarray}
g_{\mu\nu} = \eta_{ab} e^a _\mu e^b _\nu,
\label{Vierbein}
\end{eqnarray}
where $a, b, \cdots$ are local Lorentz indices taking $0, 1, 2, 3$ and $\eta_{ab} = \rm{diag} ( - 1, 1, 1, 1)$.
Now the metric condition (\ref{Gen-metric cond}) takes the form: 
\begin{eqnarray}
{\cal D}_\mu e^a _\nu \equiv D_\mu e^a _\nu + \tilde \omega^a \, _{b \mu} e^b _\nu 
- \tilde \Gamma^\rho_{\mu\nu} e^a _\rho = 0,
\label{Gen-vierbein cond}
\end{eqnarray}
where the general covariant derivative is extended to include the local Lorentz transformation whose
gauge connection is the spin connection $\tilde \omega^a \, _{b \mu}$ of weight $0$ in Weyl
geometry, and $D_\mu e^a _\nu = \partial_\mu e^a _\nu + f S_\mu e^a _\nu$ since the vierbein 
$e^a _\mu$ has weight $1$. Solving the metric condition (\ref{Gen-vierbein cond}) leads to the
expression of the spin connection in Weyl geometry:
\begin{eqnarray}
\tilde \omega_{a b \mu} = \omega_{a b \mu} + f e^c _\mu ( \eta_{ac} S_b - \eta_{bc} S_a ),
\label{spin connection}
\end{eqnarray}
where $\omega_{a b \mu}$ is the spin connection in Riemann geometry and we have defined 
$S_a \equiv e^\mu _a S_\mu$. Then, the general covariant 
derivative for a spinor field $\Psi$ of weight $- \frac{3}{2}$ reads:
\begin{eqnarray}
{\cal D}_\mu \Psi = D_\mu \Psi + \frac{i}{2} \tilde \omega_{a b \mu} S^{a b}  \Psi,
\label{spinor CD}
\end{eqnarray}
where $D_\mu \Psi = \partial_\mu \Psi - \frac{3}{2} f S_\mu \Psi$ and the Lorentz generator $S^{a b}$
for a spinor field is defined as $S^{a b} = \frac{i}{4} [ \gamma^a, \gamma^b ]$. Here we define the gamma
matrices to satisfy the Clifford algebra $\{ \gamma^a, \gamma^b \} = - 2 \eta^{ab}$.
Since the spin connection $\tilde \omega^a \, _{b \mu}$ has weight $0$, the covariant
derivative ${\cal D}_\mu \Psi$  transforms covariantly under Weyl gauge transformation:
\begin{eqnarray}
{\cal D}_\mu \Psi \rightarrow ( {\cal D}_\mu \Psi )^\prime = e^{- \frac{3}{2} \Lambda(x)} 
{\cal D}_\mu \Psi.
\label{spinor covariance}
\end{eqnarray}

Then, the Lagrangian density for a massless Dirac spinor field is of form:
\begin{eqnarray}
{\cal L} = \frac{i}{2} e \ e^\mu _a ( \bar \Psi \gamma^a {\cal D}_\mu \Psi 
- {\cal D}_\mu \bar \Psi \gamma^a \Psi ),
\label{spinor Lag}
\end{eqnarray}
where $e \equiv \sqrt{-g}, \bar \Psi \equiv \Psi^\dagger \gamma^0$, and ${\cal D}_\mu \bar \Psi$
is given by
\begin{eqnarray}
{\cal D}_\mu \bar \Psi = D_\mu \bar \Psi - \bar \Psi \frac{i}{2} \tilde \omega_{a b \mu} S^{a b}.
\label{spinor CD2}
\end{eqnarray}
Inserting Eqs. (\ref{spinor CD}) and (\ref{spinor CD2}) to the Lagrangian density (\ref{spinor Lag}), 
we find that  
\begin{eqnarray}
{\cal L} &=& \frac{i}{2} e \Bigl[ e^\mu _a  \left( \bar \Psi \gamma^a \partial_\mu \Psi 
- \partial_\mu \bar \Psi \gamma^a \Psi + \frac{i}{2} \omega_{b c \mu} \bar \Psi \{ \gamma^a,
S^{bc} \} \Psi \right)
\nonumber\\
&+& \frac{i}{2} f ( \eta_{ab} S_c - \eta_{ac} S_b ) \bar \Psi \{ \gamma^a, S^{bc} \} \Psi \Bigr].
\label{spinor Lag2}
\end{eqnarray}
The last term identically vanishes owing to the relation: 
\begin{eqnarray}
\{ \gamma^a, S^{bc} \} = - \varepsilon^{abcd} \gamma_5 \gamma_d,
\label{gamma rel}
\end{eqnarray}
where we have defined as $\gamma_5 = i \gamma^0 \gamma^1 \gamma^2 \gamma^3$ and 
$\varepsilon^{0123} = +1$. Thus, as is well known, the Weyl gauge field $S_\mu$ does not couple minimally 
to a spinor field $\Psi$.  Technically speaking, it is the absence of imaginary unit $i$ in the 
covariant derivative $D_\mu \Psi = \partial_\mu \Psi - \frac{3}{2} f S_\mu \Psi$ that induced this
decoupling of the Weyl gauge field from the spinor field. Without the imaginary unit, the terms including 
the Weyl gauge field cancel out each other in Eq. (\ref{spinor Lag}). In a similar manner, we can prove that 
the Weyl gauge field does not couple to a gauge field, i.e., the electromagnetic potential $A_\mu$ either. 
On the other hand, the Weyl gauge field can couple to a scalar field such as the Higgs field as well as a graviton.

\section{Classical theory}

We wish to consider a model of Weyl conformal gravity in Weyl geometry. It is of interest
to recall that without matter fields we have a unique classical Lagrangian which is invariant under the
Weyl gauge transformation; the Lagrangian must be of form of quadratic gravity:
\begin{eqnarray}
{\cal L}_{QG} = \sqrt{- g} \left ( - \frac{1}{2 \xi^2} \tilde C_{\mu\nu\rho\sigma} \tilde C^{\mu\nu\rho\sigma}
+ \alpha \tilde R^2 \right),
\label{L-QG}  
\end{eqnarray}
where $\xi$ and $\alpha$ are dimensionless coupling constants, and $\tilde C_{\mu\nu\rho\sigma}$ and  $\tilde R$
are a generalization of conformal tensor and scalar curvature in Weyl geometry, respectively.
Note that the Lagrangian of the Einstein-Hilbert type or the higher-derivative terms involving more than
quadratic terms are prohibited to be present by Weyl gauge symmetry. The fatal defect of the
Lagrangian (\ref{L-QG}), however, is the existence of a massless ghost which breaks unitarity in quantum regime.
Another unsatisfactory feature of the Lagrangian (\ref{L-QG}) is that it does not reduce to Einstein's
general relativity at low energies which is known to be a good description of the physics relevant to 
gravitational phenomena at such long range scales. 

Provided that we are allowed to use matter fields\footnote{As explained in the previous section, fermions
and the conventional gauge fields do not couple to the Weyl gauge field, but only the scalar field does.}, 
the situation changes and we can construct a scalar-tensor gravity of the Einstein-Hilbert type which includes at most 
the second-order derivatives of the metric tensor \cite{Dirac}:
\begin{eqnarray}
{\cal L}_{ST} = \sqrt{- g}  \, \frac{1}{2} \xi \phi^2 \tilde R,
\label{L-ST}  
\end{eqnarray}
where $\phi$ is a real scalar field.\footnote{The extension to a complex scalar field or multiple scalar
fields is straightforward.}  The most general classical Lagrangian, which is invariant under Weyl gauge
transformation and is free of the massless ghost, reads:
\begin{eqnarray}
{\cal L}_G &=& \sqrt{- g}  \biggl[  \frac{1}{2} \xi \phi^2 \tilde R - \frac{1}{4} H_{\mu\nu} H^{\mu\nu}
- \frac{1}{2} \epsilon g^{\mu\nu} D_\mu \phi D_\nu \phi - \frac{\lambda}{4 !} \phi^4
\nonumber\\
&+& \eta \left( \frac{1}{12} \phi^2 R + \frac{1}{2} g^{\mu\nu} \partial_\mu \phi \partial_\nu \phi \right) \biggr],
\label{L-G}  
\end{eqnarray}
where $\xi, \lambda, \eta$ are all dimensionless constants, and $\epsilon = \pm 1$ depending on
a normal field $\epsilon = 1$ or a ghost field $\epsilon = -1$. In this article, we limit ourselves to the case 
$6 \xi + \epsilon \neq 0$ since the specific case $6 \xi + \epsilon = 0$ leads to the same expression
as the last term with the constant $\eta$, which is called ``Weyl invariant scalar-tensor gravity'', 
when surface terms are ignored. Finally, the scalar field $\phi$ has 
the weight $-1$ so the Weyl covariant derivative in (\ref{L-G}) takes the form:\footnote{In what follows, 
we will set $f = 1$ for the coupling constant for the non-compact Abelian gauge group.}
\begin{eqnarray}
D_\mu \phi = \partial_\mu \phi - S_\mu \phi.
\label{W-covd-S}
\end{eqnarray}
 Since we have already analyzed the Weyl invariant scalar-tensor gravity in Riemann geometry \cite{Oda-W}
 and the quartic potential term has no essential role in the BRST formalism, we will put 
 $\lambda = \eta = 0$. Thus, the classical Lagrangian which is treated in this article reads:
\begin{eqnarray}
{\cal L}_c &=& \sqrt{- g}  \biggl[  \frac{1}{2} \xi \phi^2 \tilde R - \frac{1}{4} H_{\mu\nu} H^{\mu\nu}
- \frac{1}{2} \epsilon g^{\mu\nu} D_\mu \phi D_\nu \phi \biggr]
\nonumber\\
&=& \sqrt{- g}  \biggl[  \frac{1}{2} \xi \phi^2 ( R - 6 \nabla_\mu S^\mu - 6 S_\mu S^\mu ) 
- \frac{1}{4} H_{\mu\nu} H^{\mu\nu} 
\nonumber\\
&-& \frac{1}{2} \epsilon g^{\mu\nu} ( \partial_\mu \phi - S_\mu \phi ) ( \partial_\nu \phi - S_\nu \phi ) \biggr].
\label{L-c}  
\end{eqnarray}

\section{Quantum theory}

The classical Lagrangian (\ref{L-c}) is invariant under both general coordinate transformation (GCT) and 
Weyl gauge transformation. For a quantum theory we have to fix such gauge symmetries by
introducing suitable gauge-fixing conditions. After introducing the gauge-fixing conditions the quantum
Lagrangian is not longer invariant under the gauge transformations, but as residual global symmetries
the quantum Lagrangian is invariant under two BRST transformations, one of which is denoted as $\delta_B$, 
corresponding to the GCT is defined as
\begin{eqnarray}
\delta_B g_{\mu\nu} &=& - ( \nabla_\mu c_\nu+ \nabla_\nu c_\mu)
= - ( c^\alpha\partial_\alpha g_{\mu\nu} + \partial_\mu c^\alpha g_{\alpha\nu} 
+ \partial_\nu c^\alpha g_{\mu\alpha} ),
\nonumber\\
\delta_B \phi &=& - c^\lambda \partial_\lambda \phi, \quad 
\delta_B S_\mu = - c^\lambda \nabla_\lambda S_\mu - \nabla_\mu c^\lambda S_\lambda,
\nonumber\\
\delta_B c^\rho &=& - c^\lambda\partial_\lambda c^\rho, \quad
\delta_B \bar c_\rho = i B_\rho, \quad 
\delta_B B_\rho = 0, 
\label{GCT-BRST}  
\end{eqnarray}
where $c^\rho$ and $\bar c_\rho$ are respectively the Faddeev-Popov (FP) ghost and anti-ghost, 
$B_\rho$ is the Nakanishi-Lautrup (NL) field.  For convenience, in place of the NL field $B_\rho$ 
we will introduce a new NL field defined as
\begin{eqnarray}
b_\rho= B_\rho- i c^\lambda\partial_\lambda\bar c_\rho,
\label{b-rho-field}  
\end{eqnarray}
and its BRST transformation reads:
\begin{eqnarray}
\delta_B b_\rho= - c^\lambda\partial_\lambda b_\rho.
\label{b-BRST}  
\end{eqnarray}

The other BRST transformation, which is denoted as $\bar \delta_B$, 
corresponding to the Weyl transformation is defined as
\begin{eqnarray}
\bar \delta_B g_{\mu\nu} &=& 2 c g_{\mu\nu}, \quad
\bar \delta_B \phi = - c \phi, \quad 
\bar \delta_B S_\mu = - \partial_\mu c,  
\nonumber\\
\bar \delta_B \bar c &=& i B, \quad 
\bar \delta_B c = \bar \delta_B B = 0, 
\label{Weyl-BRST}  
\end{eqnarray}
where $c$ and $\bar c$ are respectively the FP ghost and FP anti-ghost, 
$B$ is the NL field. Note that the two BRST transformations are nilpotent, i.e.,
\begin{eqnarray}
\delta_B^2 = \bar \delta_B^2 = 0.   
\label{Nilpotent}  
\end{eqnarray}

To complete the two BRST transformations, we have to fix not only the GCT BRST transformation
$\delta_B$ on $c, \bar c$ and $B$ but also the Weyl BRST transformation $\delta_B$ on
$c^\rho, \bar c_\rho$ and $b_\rho$. The BRST transformations on these fields are
fixed by requiring that the two BRST transformations anti-commute with each other, that is, \cite{Oda-W}  
\begin{eqnarray}
\{ \delta_B, \bar \delta_B \} \equiv \delta_B \bar \delta_B + \bar \delta_B \delta_B = 0.
\label{GCT-Weyl-BRST}  
\end{eqnarray}
Then, the resultant BRST transformations take the form:
\begin{eqnarray}
\delta_B B &=& - c^\lambda\partial_\lambda B, \quad
\delta_B c = - c^\lambda\partial_\lambda c, \quad
\delta_B \bar c = - c^\lambda\partial_\lambda \bar c,
\nonumber\\
\bar \delta_B b_\rho &=& \bar \delta_B c^\rho = \bar \delta_B \bar c_\rho = 0.   
\label{BRST2}  
\end{eqnarray}

In this context, it is worthwhile to recall that the gauge condition for the GCT must be invariant 
under Weyl gauge transformation while the one for Weyl transformation must be invariant
under GCT in order for the two BRST transformations to anti-commute.
In that case we can consider the two BRST transformations separately.  The suitable
gauge condition for the GCT is almost unique and is called ``the extended de Donder 
gauge'' \cite{Oda-W}:\footnote{Let us note that this gauge condition breaks the general coordinate
invariance, but it is invariant under the general linear transformation $GL(4)$. Thus, the quantum Lagrangian
which is obtained shortly is also invaraint under the $GL(4)$.}
\begin{eqnarray}
\partial_\mu ( \tilde g^{\mu\nu} \phi^2 ) = 0,
\label{Ext-de-Donder}  
\end{eqnarray}
where we have defined $\tilde g^{\mu\nu} \equiv \sqrt{-g} g^{\mu\nu}$.

On the other hand, we have a few candidates for the gauge-fixing condition for the Weyl transformation, 
which must be invariant under the GCT, i.e., a scalar quantity. The first one is the well-known
``unitary gauge'', $\phi = \rm{constant}$, which is taken to show that Weyl invariant 
scalar-tensor gravity is equivalent to the Einstein-Hilbert term. The other
gauge condition is the Lorenz gauge, $\nabla_\mu S^\mu = 0$, which is usually adopted 
in quantum field theories. However, it turns out that these gauge conditions are not so
interesting in the present context since they do not allow for conformal symmetry to remain.
Hence, we shall choose, what we call, ``the scalar gauge condition'' \cite{Oda-W}:
\begin{eqnarray}
\partial_\mu ( \tilde g^{\mu\nu} \phi \partial_\nu \phi ) = 0,
\label{Scalar-gauge}  
\end{eqnarray}
which can be alternatively written as 
\begin{eqnarray}
\Box \, \phi^2 = 0.
\label{Alt-Scalar-gauge}  
\end{eqnarray}

After taking the extended de Donder gauge condition (\ref{Ext-de-Donder}) for the GCT and the scalar gauge condition
(\ref{Scalar-gauge}) for the Weyl transformation, the gauge-fixed and BRST invariant quantum Lagrangian is given by
\begin{eqnarray}
{\cal L}_q &=& {\cal L}_c + {\cal L}_{GF+FP} + \bar {\cal L}_{GF+FP}
\nonumber\\
&=& {\cal L}_c + i \delta_B ( \tilde g^{\mu\nu} \phi^2 \partial_\mu \bar c_\nu )
+ i \bar \delta_B \left[ \bar c \partial_\mu ( \tilde g^{\mu\nu} \phi \partial_\nu \phi ) \right] 
\nonumber\\
&=& \sqrt{- g}  \biggl[  \frac{1}{2} \xi \phi^2 ( R - 6 \nabla_\mu S^\mu - 6 S_\mu S^\mu ) 
- \frac{1}{4} H_{\mu\nu} H^{\mu\nu} 
- \frac{1}{2} \epsilon g^{\mu\nu} D_\mu \phi D_\nu \phi \biggr]
\nonumber\\
&-& \tilde g^{\mu\nu} \phi^2 ( \partial_\mu b_\nu + i \partial_\mu \bar c_\lambda  \partial_\nu c^\lambda )
+ \tilde g^{\mu\nu} \phi \partial_\mu B \partial_\nu \phi - i \tilde g^{\mu\nu} \phi^2 \partial_\mu \bar c 
\partial_\nu c,
\label{ST-q-Lag}  
\end{eqnarray}
where surface terms are dropped. 

From the Lagrangian ${\cal L}_q$, it is straightforward to derive the field equations by taking 
the variation with respect to $g_{\mu\nu}, S_\mu, \phi, b_\nu, B, c^\rho, \bar c_\rho, c$ and $\bar c$ in order:
\begin{eqnarray}
&{}& \frac{1}{2} \xi \phi^2 G_{\mu\nu} - \frac{1}{2} \xi ( \nabla_\mu \nabla_\nu - g_{\mu\nu} \Box ) \phi^2  
- 3 \xi \phi^2 ( S_\mu S_\nu - \frac{1}{2} g_{\mu\nu} S_\alpha S^\alpha )
\nonumber\\
&{}& + 3 \xi \phi ( S_\mu \partial_\nu \phi + S_\nu \partial_\mu \phi - g_{\mu\nu} S^\alpha \partial_\alpha \phi )
- \frac{1}{2} H_{\mu\alpha} H_\nu \, ^\alpha + \frac{1}{8} g_{\mu\nu} H_{\alpha\beta}^2 
\nonumber\\
&{}& - \frac{1}{2} \epsilon \left[ D_\mu \phi D_\nu \phi - \frac{1}{2} g_{\mu\nu} ( D_\alpha \phi )^2 \right] 
- \frac{1}{2} ( E_{\mu\nu} - \frac{1}{2} g_{\mu\nu} E ) = 0, 
\nonumber\\
&{}& ( 6 \xi + \epsilon ) g^{\mu\nu} \phi D_\nu \phi - \nabla_\nu H^{\mu\nu} = 0,
\nonumber\\
&{}& \xi \phi^2 ( R - 6 \nabla_\mu S^\mu - 6 S_\mu S^\mu ) + \epsilon \frac{1}{\sqrt{-g}} 
\phi D_\mu ( \tilde g^{\mu\nu} \phi D_\nu \phi ) - E 
\nonumber\\
&{}& - 2 g^{\mu\nu} \phi \partial_\mu B \partial_\nu \phi 
- \phi^2 \Box B = 0,
\nonumber\\
&{}& \partial_\mu ( \tilde g^{\mu\nu} \phi^2 ) = 0, \qquad
\partial_\mu ( \tilde g^{\mu\nu} \phi \partial_\nu \phi ) = 0, 
\nonumber\\
&{}& g^{\mu\nu} \partial_\mu \partial_\nu\bar c_\rho = g^{\mu\nu} \partial_\mu \partial_\nu c^\rho 
= g^{\mu\nu} \partial_\mu \partial_\nu\bar c = g^{\mu\nu} \partial_\mu \partial_\nu c = 0.
\label{q-field-eq}  
\end{eqnarray}
where $G_{\mu\nu} \equiv R_{\mu\nu} - \frac{1}{2} g_{\mu\nu} R$ denotes the Einstein tensor,
while $E_{\mu\nu}$ and $E$ are defined as
\begin{eqnarray}
E_{\mu\nu} &=& \phi^2 ( \partial_\mu b_\nu 
+ i \partial_\mu \bar c_\lambda  \partial_\nu c^\lambda )
- \phi \partial_\mu B \partial_\nu \phi + i \phi^2 \partial_\mu \bar c \partial_\nu c
+ ( \mu \leftrightarrow \nu ),
\nonumber\\
E &=& g^{\mu\nu} E_{\mu\nu}.
\label{E}  
\end{eqnarray}
Moreover, since $\tilde g^{\mu\nu} D_\nu \phi$ has the weight $1$, the Weyl covariant derivative 
is defined as
\begin{eqnarray}
D_\mu ( \tilde g^{\mu\nu} D_\nu \phi ) = \partial_\mu ( \tilde g^{\mu\nu} D_\nu \phi ) 
+ S_\mu \tilde g^{\mu\nu} D_\nu \phi.
\label{W-cov-d}  
\end{eqnarray}

When we introduce the dilaton $\sigma (x)$ by defining
\begin{eqnarray}
\phi (x) \equiv e^{\sigma (x)},
\label{Dilaton}  
\end{eqnarray}
the two gauge-fixing conditions in (\ref{q-field-eq}), or equivalently, Eqs. (\ref{Ext-de-Donder}) and
(\ref{Scalar-gauge}) lead to a very simple d'Alembert-like equation for the dilaton:
\begin{eqnarray}
g^{\mu\nu} \partial_\mu \partial_\nu \sigma = 0.
\label{Dilaton-eq}  
\end{eqnarray}
It is worthwhile to notice that it is not the scalar field $\phi$ but the dilaton $\sigma$ that satisfies 
this type of equation. 

In order to show that the auxiliary field $B$ also obeys the same type of equation, let us take 
account of the trace part of the Einstein equation, i.e., the first field equation 
in (\ref{q-field-eq}), which gives us the equation:
\begin{eqnarray}
\xi \phi^2 R - 6 \xi \phi^2 S_\alpha S^\alpha + 12 \xi \phi S^\alpha \partial_\alpha \phi 
- \epsilon ( D_\alpha \phi )^2 -  E = 0. 
\label{Trace-E-eq}  
\end{eqnarray}
Next, we can rewrite the field equation for $\phi$, the third equation in (\ref{q-field-eq}), as
\begin{eqnarray}
&{}& \xi \phi^2 R - 6 \xi \phi^2 S_\alpha S^\alpha + 12 \xi \phi S^\alpha \partial_\alpha \phi 
- \epsilon ( D_\alpha \phi )^2 -  E 
\nonumber\\
&{}& - ( 6 \xi + \epsilon ) \phi^2 g^{\mu\nu} \partial_\mu S_\nu 
- 2 g^{\mu\nu} \phi \partial_\mu B \partial_\nu \phi - \phi^2 \Box B = 0. 
\label{Phi-eq}  
\end{eqnarray}
Using Eqs. (\ref{Trace-E-eq}) and (\ref{Phi-eq}), we can obtain the equation:
\begin{eqnarray}
g^{\mu\nu} \partial_\mu \partial_\nu B + ( 6 \xi + \epsilon ) g^{\mu\nu} \partial_\mu S_\nu = 0. 
\label{Trace&Phi-eq}  
\end{eqnarray}
Now we are ready to prove 
\begin{eqnarray}
g^{\mu\nu} \partial_\mu S_\nu = 0. 
\label{S-eq}  
\end{eqnarray}
To do that, let us consider the field equation for $S_\mu$ in (\ref{q-field-eq}), multiply by
$\sqrt{-g}$, and then operate the covariant derivative consequently leading to:
\begin{eqnarray}
\sqrt{-g} \nabla_\mu \nabla_\nu H^{\mu\nu}  = ( 6 \xi + \epsilon ) \nabla_\mu 
( \tilde g^{\mu\nu} \phi D_\nu \phi ).
\label{d-S-eq}  
\end{eqnarray}
The LHS of Eq. (\ref{d-S-eq}) is identically zero and $6 \xi + \epsilon \neq 0$
by our assumption, we find that 
\begin{eqnarray}
\nabla_\mu ( \tilde g^{\mu\nu} \phi D_\nu \phi ) = 0.
\label{d-S-eq2}  
\end{eqnarray}
Using the formula: 
\begin{eqnarray}
\nabla_\mu ( \tilde g^{\mu\nu} A_\nu ) = \partial_\mu ( \tilde g^{\mu\nu} A_\nu ),
\label{Math-formula}  
\end{eqnarray}
which holds for an arbitrary covariant vector $A_\mu$, Eq. (\ref{d-S-eq2}) is reduced
to the form:
\begin{eqnarray}
\partial_\mu ( \tilde g^{\mu\nu} \phi \partial_\nu \phi - \tilde g^{\mu\nu} \phi^2 S_\nu ) = 0.
\label{d-S-eq3}  
\end{eqnarray}
Then, using the gauge conditions (\ref{Ext-de-Donder}) and (\ref{Scalar-gauge}),
we can reach the equation (\ref{S-eq}). Hence, Eq. (\ref{Trace&Phi-eq}) implies that the auxiliary field $B$
obeys the equation:
\begin{eqnarray}
g^{\mu\nu} \partial_\mu \partial_\nu B = 0.
\label{B-eq}  
\end{eqnarray}

Surprisingly enough, using the Weyl BRST transformation, we can show this equation (\ref{B-eq}) 
in the simplest way. For this aim, let us start with the field equation for $\bar c$ in (\ref{q-field-eq}):
\begin{eqnarray}
g^{\mu\nu} \partial_\mu \partial_\nu \bar c = 0.
\label{Bar-c-eq}  
\end{eqnarray}
Operating $\bar \delta_B$ on this equation leads to  
\begin{eqnarray}
- 2 c g^{\mu\nu} \partial_\mu \partial_\nu \bar c + i g^{\mu\nu} \partial_\mu \partial_\nu B = 0.
\label{Bar-c-eq2}  
\end{eqnarray}
The first term on the LHS is vanishing owing to (\ref{Bar-c-eq}), so we can arrive at the
equation (\ref{B-eq}).

In a perfectly similar manner, we can show that the Nakanishi-Lautrup auxiliary field $b_\rho$ 
satisfies the d'Alembert-like equation by either an explicit calculation or using the BRST 
transformation for the GCT. Here we present only the latter proof since the former one
was given in our previous paper \cite{Oda-Q}.

Let us start with the field equation for $\bar c_\rho$ in (\ref{q-field-eq}):
\begin{eqnarray}
g^{\mu\nu} \partial_\mu \partial_\nu \bar c_\rho = 0.
\label{Bar-c-rho-eq}  
\end{eqnarray}
Taking the GCT BRST transformation of this equation yields:
\begin{eqnarray}
( - \partial_\lambda g^{\mu\nu} c^\lambda + g^{\mu\alpha} \partial_\alpha c^\nu 
+ g^{\nu\alpha} \partial_\alpha c^\mu )  \partial_\mu \partial_\nu \bar c_\rho 
+ i g^{\mu\nu} \partial_\mu \partial_\nu B_\rho = 0,
\label{Bar-c-rho-eq2}  
\end{eqnarray}
where we have used the GCT BRST transformation (\ref{GCT-BRST}). Substituting the definition
of $b_\rho$ in Eq. (\ref{b-rho-field}) into (\ref{Bar-c-rho-eq2}), we have the equation for $b_\rho$:
\begin{eqnarray}
i g^{\mu\nu} \partial_\mu \partial_\nu b_\rho = g^{\mu\nu} \partial_\mu \partial_\nu ( c^\lambda 
\partial_\lambda \bar c_\rho ) - ( - \partial_\lambda g^{\mu\nu} c^\lambda 
+ 2 g^{\mu\alpha} \partial_\alpha c^\nu )  \partial_\mu \partial_\nu \bar c_\rho.
\label{Bar-c-rho-eq3}  
\end{eqnarray}
With the help of Eq. (\ref{Bar-c-rho-eq}) and the field equation for $c^\rho$ in (\ref{q-field-eq}), 
the RHS is found to be vanishing so we have the desired equation:
\begin{eqnarray}
g^{\mu\nu} \partial_\mu \partial_\nu b_\rho = 0.
\label{b-rho-eq}  
\end{eqnarray}
In other words, setting $X^M = \{ x^\mu, b_\mu, \sigma, B, c^\mu, \bar c_\mu, c, \bar c \}$, $X^M$
turns out to obey the very simple equation:
\begin{eqnarray}
g^{\mu\nu} \partial_\mu \partial_\nu X^M = 0.
\label{X-M-eq}  
\end{eqnarray}
This fact, together with the gauge condition $\partial_\mu ( \tilde g^{\mu\nu} \phi^2 ) = 0$
produces the two kinds of conserved currents:
\begin{eqnarray}
{\cal P}^{\mu M} &\equiv& \tilde g^{\mu\nu} \phi^2 \partial_\nu X^M 
= \tilde g^{\mu\nu} \phi^2 \bigl( 1 \overset{\leftrightarrow}{\partial}_\nu X^M \bigr)
\nonumber\\
{\cal M}^{\mu M N} &\equiv& \tilde g^{\mu\nu} \phi^2 \bigl( X^M 
\overset{\leftrightarrow}{\partial}_\nu Y^N \bigr),
\label{Cons-currents}  
\end{eqnarray}
where we have defined $X^M \overset{\leftrightarrow}{\partial}_\mu Y^N \equiv
X^M \partial_\mu Y^N - ( \partial_\mu X^M ) Y^N$. These conserved currents 
constitute a Poincar${\rm{\acute{e}}}$-like $\IOSp(10|10)$ supersymmetry as will be
shown later.

\section{Canonical quantization and equal-time commutation relations}

In this section, after introducing the Poisson brackets, we will evaluate various equal-time 
commutation relations (ETCRs) among fundamental variables. To simplify various expressions, we will obey 
the following abbreviations adopted in the textbook of Nakanishi and Ojima \cite{N-O-text}:
\begin{eqnarray}
[ A, B^\prime ] &=& [ A(x), B(x^\prime) ] |_{x^0 = x^{\prime 0}},
\qquad \delta^3 = \delta(\vec{x} - \vec{x}^\prime), 
\nonumber\\
\tilde f &=& \frac{1}{\tilde g^{00}} = \frac{1}{\sqrt{-g} g^{00}},
\label{abbreviation}  
\end{eqnarray}
where we assume that $\tilde g^{00}$ is invertible.  Here the above brackets $[ A, B^\prime ]$
symbolically describe the Poisson brackets and the ETCRs.

First of all, let us set up the Poisson brackets of canonical variables: 
\begin{eqnarray}
&{}& \{ g_{\mu\nu}, \pi_g^{\rho\lambda\prime} \}_P =  \frac{1}{2} ( \delta_\mu^\rho\delta_\nu^\lambda 
+ \delta_\mu^\lambda\delta_\nu^\rho) \delta^3,  \quad 
\{ \phi, \pi_\phi^\prime \}_P =  \delta^3,  \quad
\{ S_\mu, \pi_S^{\nu\prime} \}_P = \delta_\mu^\nu \delta^3,
\nonumber\\
&{}& \{ c^\sigma, \pi_{c \lambda}^\prime \}_P = \{ \bar c_\lambda, \pi_{\bar c}^{\sigma\prime} \}_P
= \delta_\lambda^\sigma \delta^3,  \quad
\{ B, \pi_B^\prime \}_P = \{ c, \pi_c^\prime \}_P 
\nonumber\\
&{}& = \{ \bar c, \pi_{\bar c}^\prime \}_P = \delta^3,
\label{CCRs}  
\end{eqnarray}
where the other Poisson brackets vanish.
Here the canonical variables are $g_{\mu\nu}, \phi, S_\mu, B, c^\rho, \bar c_\rho, c, \bar c$ and the corresponding canonical
conjugate momenta are $\pi_g^{\mu\nu}, \pi_\phi, \pi_S^\mu, \pi_B, \pi_{c \rho}, \pi_{\bar c}^\rho, \pi_c, \pi_{\bar c}$, respectively 
and the $b_\mu$ field is regarded as not a canonical variable but a conjugate momentum of $\tilde g^{0 \mu}$. 

To remove second order derivatives of the metric involved in $R$, we perform the integration by parts once and
rewrite the Lagrangian (\ref{ST-q-Lag}) as
\begin{eqnarray}
{\cal L}_q &=& - \frac{1}{2} \xi \tilde g^{\mu\nu} \phi^2 ( \Gamma^\sigma_{\mu\nu} \Gamma^\alpha_{\sigma\alpha}  
-  \Gamma^\sigma_{\mu\alpha} \Gamma^\alpha_{\sigma\nu} + 6 S_\mu S_\nu ) 
- \xi \phi \partial_\mu \phi ( \tilde g^{\alpha\beta} \Gamma^\mu_{\alpha\beta}  
- \tilde g^{\mu\nu} \Gamma^\alpha_{\nu\alpha} ) 
\nonumber\\
&+& 6 \xi \tilde g^{\mu\nu} \phi S_\mu \partial_\nu \phi - \frac{1}{4} \sqrt{-g} H_{\mu\nu} H^{\mu\nu}
- \frac{1}{2} \epsilon \tilde g^{\mu\nu} D_\mu \phi D_\nu \phi
+ \partial_\mu ( \tilde g^{\mu\nu} \phi^2 ) b_\nu
\nonumber\\ 
&-& i \tilde g^{\mu\nu} \phi^2 \partial_\mu \bar c_\rho \partial_\nu c^\rho 
+ \tilde g^{\mu\nu} \partial_\mu B \phi \partial_\nu \phi
- i \tilde g^{\mu\nu} \phi^2 \partial_\mu \bar c \partial_\nu c
+ \partial_\mu {\cal{V}}^\mu,
\label{Mod-ST-q-Lag}  
\end{eqnarray}
where we have also integrated by parts two terms with the linear $S_\mu$ and $b_\mu$, and a surface term 
${\cal{V}}^\mu$ is thus given by
\begin{eqnarray}
{\cal{V}}^\mu =  \frac{1}{2} \xi \phi^2 ( \tilde g^{\alpha\beta} \Gamma^\mu_{\alpha\beta} 
- \tilde g^{\mu\nu} \Gamma^\alpha_{\nu\alpha} ) - 3 \xi \tilde g^{\mu\nu} \phi^2 S_\nu
- \tilde g^{\mu\nu} \phi^2 b_\nu.
\label{surface}  
\end{eqnarray}
Using this Lagrangian, the concrete expressions for canonical conjugate momenta become:
\begin{eqnarray}
\pi_g^{\mu\nu} &=& \frac{\partial {\cal L}_q}{\partial \dot g_{\mu\nu}} 
\nonumber\\
&=& - \frac{1}{4} \sqrt{-g} \xi \phi^2 \Bigl[ - g^{0 \lambda} g^{\mu\nu} g^{\sigma\tau} 
- g^{0 \tau} g^{\mu\lambda} g^{\nu\sigma}
- g^{0 \sigma} g^{\mu\tau} g^{\nu\lambda} + g^{0 \lambda} g^{\mu\tau} g^{\nu\sigma} 
\nonumber\\
&+& g^{0 \tau} g^{\mu\nu} g^{\lambda\sigma}
+ \frac{1}{2} ( g^{0 \mu} g^{\nu\lambda} + g^{0 \nu} g^{\mu\lambda} ) g^{\sigma\tau} \Bigr] \partial_\lambda g_{\sigma\tau}
\nonumber\\
&-& \sqrt{-g} \Bigl[ \frac{1}{2} ( g^{0 \mu} g^{\rho\nu} + g^{0 \nu} g^{\rho\mu} ) 
- g^{\mu\nu} g^{\rho 0} \Bigr] \xi \phi \partial_\rho \phi 
\nonumber\\
&-& \frac{1}{2} \sqrt{-g} ( g^{0 \mu} g^{\nu\rho} + g^{0 \nu} g^{\mu\rho} - g^{0 \rho} g^{\mu\nu} ) \phi^2 b_\rho,
\nonumber\\
\pi_\phi &=& \frac{\partial {\cal L}_q}{\partial \dot \phi} = - \epsilon \tilde g^{0 \mu} D_\mu \phi
+ 2 \tilde g^{0 \mu} \phi b_\mu + \xi \phi ( - \tilde g^{\alpha\beta} \Gamma^0_{\alpha\beta} 
+ \tilde g^{0 \alpha} \Gamma^\beta_{\alpha\beta} ) 
\nonumber\\
&+& 6 \xi \tilde g^{0 \mu} \phi S_\mu + \tilde g^{0 \mu} \phi \partial_\mu B,
\nonumber\\
\pi_S^\mu &=& \frac{\partial {\cal L}_q}{\partial \dot S_\mu} = - \sqrt{-g} H^{0\mu},  \quad
\pi_B = \frac{\partial {\cal L}_q}{\partial \dot B} = \tilde g^{0 \mu} \phi \partial_\mu \phi,
\nonumber\\
\pi_{c \sigma} &=& \frac{\partial {\cal L}_q}{\partial \dot c^\sigma} = - i \tilde g^{0 \mu} \phi^2 \partial_\mu \bar c_\sigma, \quad
\pi_{\bar c}^\sigma = \frac{\partial {\cal L}_q}{\partial \dot {\bar c}_\sigma} = i \tilde g^{0 \mu} \phi^2 \partial_\mu c^\sigma,
\nonumber\\
\pi_c &=& \frac{\partial {\cal L}_q}{\partial \dot c} = - i \tilde g^{0 \mu} \phi^2 \partial_\mu \bar c, \quad
\pi_{\bar c} = \frac{\partial {\cal L}_q}{\partial \dot {\bar c}} = i \tilde g^{0 \mu} \phi^2 \partial_\mu c,
\label{CCM}  
\end{eqnarray}
where we have defined the time derivative such as $\dot g_{\mu\nu} \equiv \frac{\partial g_{\mu\nu}}{\partial t}
\equiv \partial_0 g_{\mu\nu}$, and differentiation of ghosts is taken from the right. 

It can be easily seen that we have a primary constraint:
\begin{eqnarray}
\Psi_1 \equiv \pi_S^0 \approx 0.
\label{Primary}  
\end{eqnarray}
Let us recall that a secondary constraint comes from the consistency under time evolution of the primary contraint:
\begin{eqnarray}
\Psi_2 \equiv \dot \pi_S^0 = \{ \pi_S^0, H_T \}_P \approx 0,
\label{Second1}  
\end{eqnarray}
where $H_T$ is the Hamiltonian of the system at hand, which is defined as
\begin{eqnarray}
H_T &\equiv& \int d^3 x \, {\cal{H}}_T
\nonumber\\
&=& \int d^3 x \, (  \pi_g^{\mu\nu} \dot g_{\mu\nu} + \pi_\phi \dot \phi + \pi_S^\mu \dot S_\mu
+ \pi_B \dot B + \pi_{c \mu} \dot c^\mu + \pi_{\bar c}^\mu \dot {\bar c}_\mu 
\nonumber\\
&+& \pi_c \dot c + \pi_{\bar c} \dot{\bar c} - {\cal L}_q ).
\label{Hamil}  
\end{eqnarray}

In order to obtain the Hamiltonian, we have to express the time derivatives of the canonical variables in terms of
the canonical conjugate momenta in (\ref{CCM}). To do that, let us first consider $\pi_B$, which gives us the expression 
of $\dot \phi$ as
\begin{eqnarray}
\dot \phi = \tilde f \left( \frac{1}{\phi} \pi_B - \tilde g^{0i} \partial_i \phi \right).
\label{dot-phi}  
\end{eqnarray}
Next, let us turn our attention to the $(kl)$-components of $\pi_g^{\mu\nu}$, which take the form:
\begin{eqnarray}
\pi_g^{kl} = \hat A^{kl} + \hat B^{kl \rho} b_\rho + \hat C^{klmn} \dot g_{mn} + \hat D^{kl} \dot \phi,
\label{dot-pi-kl}  
\end{eqnarray}
where $\hat A^{kl}, \hat B^{kl \rho}, \hat C^{klmn}$ and $\hat D^{kl}$ commute with $g_{mn}$ and are defined as
\begin{eqnarray}
\hat A^{kl} &=& -\frac{1}{4} \sqrt{-g} \phi^2 \Bigl[ - g^{0m} g^{kl} g^{\sigma\tau} - g^{0\tau} g^{km} g^{l\sigma}
- g^{0\sigma} g^{k\tau} g^{lm} + g^{0m} g^{k\tau} g^{l\sigma} + g^{0\tau} g^{kl} g^{m\sigma} 
\nonumber\\
&+& \frac{1}{2} ( g^{0k} g^{lm} + g^{0l} g^{km} ) g^{\sigma\tau} \Bigr] \partial_m g_{\sigma\tau}
- \sqrt{-g} \xi \phi \left[ \frac{1}{2} ( g^{0k} g^{lm} + g^{0l} g^{km} ) - g^{kl} g^{0m} \right] \partial_m \phi,
\nonumber\\
\hat B^{kl \rho} &=& -\frac{1}{2} \sqrt{-g} \phi^2 ( g^{0k} g^{l\rho} + g^{0l} g^{k\rho}  - g^{0\rho} g^{kl} ),
\nonumber\\
\hat C^{klmn} &=& -\frac{1}{4} \sqrt{-g} \xi \phi^2 ( - g^{00} g^{kl} g^{mn} - g^{0n} g^{0k} g^{lm}
- g^{0m} g^{kn} g^{0l} + g^{00} g^{kn} g^{lm} 
\nonumber\\
&+& g^{0n} g^{kl} g^{0m} + g^{0k} g^{0l} g^{mn} ),
\nonumber\\
\hat D^{kl} &=& \sqrt{-g} \xi \phi ( g^{00} g^{kl} - g^{0k} g^{0l} ). 
\label{hat-ABCD}  
\end{eqnarray}
Solving (\ref{dot-pi-kl}) with respect to $\dot g_{kl}$ together with Eq. (\ref{dot-phi}) leads to: 
\begin{eqnarray}
\dot g_{kl} = \hat C_{klmn}^{-1} \left[ \pi_g^{mn} - \hat A^{mn} - \hat B^{mn \rho} b_\rho 
- \hat D^{mn} \tilde f \left( \frac{1}{\phi} \pi_B - \tilde g^{0i} \partial_i \phi \right) \right],
\label{dot-g-kl}  
\end{eqnarray}
where $\hat C_{klmn}^{-1}$ is the inverse matrix of $\hat C^{klmn}$ given by
\begin{eqnarray}
\hat C_{klmn}^{-1} &=& \frac{2}{\xi \phi^2} \tilde f ( g_{kl} g_{mn} - g_{km} g_{ln} - g_{kn} g_{lm} ),
\nonumber\\
\hat C^{klmn} \hat C_{mnij}^{-1} &=& \frac{1}{2} ( \delta_i^k \delta_j^l + \delta_i^l \delta_j^k ).
\label{Inv-C}  
\end{eqnarray}
Using the extended de Donder gauge condition (\ref{Ext-de-Donder}), $\dot g_{00}$ and
$\dot g_{0k}$ are described as
\begin{eqnarray}
\dot g_{00} &=& \frac{1}{g^{00}} \left( g^{ij} \dot g_{ij} - 2 g^{\alpha i} \partial_i g_{0 \alpha} 
+ \frac{4}{\phi} \dot \phi \right)
\nonumber\\
&=& \frac{1}{g^{00}} \Biggl\{ g^{kl} \hat C_{klmn}^{-1} \left[ \pi_g^{mn} - \hat A^{mn} 
- \hat B^{mn \rho} b_\rho - \hat D^{mn} \tilde f \left( \frac{1}{\phi} \pi_B 
- \tilde g^{0i} \partial_i \phi \right) \right]
\nonumber\\
&-& 2 g^{\alpha i} \partial_i g_{0 \alpha} + \frac{4}{\phi} \tilde f \left( \frac{1}{\phi} \pi_B 
- \tilde g^{0i} \partial_i \phi \right) \Biggr\},
\nonumber\\
\dot g_{0k} &=& \frac{1}{g^{00}} \left( - g^{0j} \dot g_{jk} - g^{\alpha i} \partial_i g_{\alpha k} 
+ \frac{1}{2} g^{\alpha\beta} \partial_k g_{\alpha\beta} + \frac{2}{\phi} \partial_k \phi \right)
\nonumber\\
&=& \frac{1}{g^{00}} \Biggl\{ - g^{0j} \hat C_{jkmn}^{-1} \left[ \pi_g^{mn} - \hat A^{mn} 
- \hat B^{mn \rho} b_\rho - \hat D^{mn} \tilde f \left( \frac{1}{\phi} \pi_B 
- \tilde g^{0i} \partial_i \phi \right) \right]
\nonumber\\
&-& g^{\alpha i} \partial_i g_{\alpha k} + \frac{1}{2} g^{\alpha\beta} \partial_k g_{\alpha\beta} 
+ \frac{2}{\phi} \partial_k \phi \Biggr\}.
\label{dot-g-0mu}  
\end{eqnarray}

In a similar manner, based on $\pi_\phi, \pi_S^\mu, \pi_{c \sigma}, \pi_{\bar c}^\sigma, \pi_c$ and 
$\pi_{\bar c}$ in Eq. (\ref{CCM}), the time derivatives $\dot B, \dot S_k, \dot S_0, \dot{\bar c}_\sigma,
\dot c^\sigma, \dot{\bar c}$ and $\dot c$ can be expressed in terms of the canonical conjugate momenta
as follows:
\begin{eqnarray}
\dot B &=& \tilde f \frac{1}{\phi} \Bigl[ \pi_\phi + \epsilon \frac{1}{\phi} \pi_B - ( 6 \xi + \epsilon )
\tilde g^{0\mu} \phi S_\mu - 2 \tilde g^{0\mu} \phi b_\mu - \tilde g^{0i} \phi \partial_i B \Bigr] 
\nonumber\\
&-& \xi \tilde f \Biggl\{ ( \tilde g^{00} g^{ij} - \tilde g^{0i} g^{0j} ) \hat C_{ijmn}^{-1} \Bigl[ \pi_g^{mn} 
- \hat A^{mn} - \hat B^{mn \rho} b_\rho - \xi \tilde f ( \tilde g^{00} g^{mn} 
\nonumber\\
&-& \tilde g^{0m} g^{0n} ) ( \pi_B - \tilde g^{0k} \phi \partial_k \phi ) \Bigr]
+ ( \tilde g^{0i} g^{\alpha\beta} - \tilde g^{0\alpha} g^{i\beta} ) \partial_i g_{\alpha\beta}  \Biggr\},
\nonumber\\
\dot S_k &=& \partial_k S_0 + \tilde f ( - g_{kj} \pi_S^j + \tilde g^{0j} H_{kj} ),
\nonumber\\
\dot S_0 &=& - \tilde f \left\{ \tilde g^{0i} \left[ 2 \partial_i S_0 + \tilde f ( - g_{ij} \pi_S^j 
+ \tilde g^{0j} H_{ij} ) \right] + \tilde g^{ij} \partial_i S_j \right\},
\nonumber\\
\dot {\bar c}_\sigma &=& i \tilde f \phi^{-2} \pi_{c \sigma} - \tilde f \tilde g^{0i} \partial_i \bar c_\sigma,
\nonumber\\
\dot c^\sigma &=& - i \tilde f \phi^{-2} \pi_{\bar c}^\sigma - \tilde f \tilde g^{0i} \partial_i c^\sigma,
\nonumber\\
\dot {\bar c} &=& i \tilde f \phi^{-2} \pi_c - \tilde f \tilde g^{0i} \partial_i \bar c,
\nonumber\\
\dot c &=& - i \tilde f \phi^{-2} \pi_{\bar c} - \tilde f \tilde g^{0i} \partial_i c,
\label{dot-many}  
\end{eqnarray}
where we have used Eq. (\ref{S-eq}) in deriving $\dot S_0$.

Finally, we can also express the $b_\mu$ field in terms of canonical conjugate momenta. Since the $b_\mu$ field
is regarded as a conjugate momentum of $\tilde g^{0\mu}$, we begin with $\pi_g^{\alpha0}$ which
has a structure:
\begin{eqnarray}
\pi_g^{\alpha0} = A^\alpha + B^{\alpha\beta} \partial_\beta \phi + C^{\alpha\beta} b_\beta,
\label{pi-0alpha}  
\end{eqnarray}
where $A^\alpha, B^{\alpha\beta}$ and $C^{\alpha\beta} = - \frac{1}{2} \tilde g^{00} g^{\alpha\beta} \phi^2$
do not include $\dot g_{\mu\nu}$, and $B^{\alpha\beta} \partial_\beta \phi$ does not have $\dot \phi$.
Solving this equation with respect to $b_\mu$ leads to: 
\begin{eqnarray}
b_\mu &=& - 2 \tilde f \phi^{-2} g_{\mu\alpha} \pi_g^{\alpha0} - \frac{1}{2} \xi \tilde f
\biggl[ \delta_\mu^0 \left( \tilde g^{0\tau} g^{\lambda\sigma} -  \frac{1}{2} \tilde g^{0\lambda} 
g^{\sigma\tau} \right) \partial_\lambda g_{\sigma\tau} 
\nonumber\\
&-& \left( \tilde g^{0\tau} g^{0\sigma} -  \frac{1}{2} \tilde g^{00} g^{\sigma\tau} \right) 
\partial_\mu g_{\sigma\tau} \biggr]
- \xi \phi^{-1} \left( \partial_\mu \phi - \delta_\mu^0 \tilde f \tilde g^{0\alpha} \partial_\alpha \phi \right).
\label{b-mu}  
\end{eqnarray}
Note that the RHS of this equation does not involve $\dot g_{\mu\nu}$ and $\dot \phi$ as can be verified explicitly.
Incidentally, the relation (\ref{pi-0alpha}) is utilized to derive some useful Poisson brackets such as
$\{ g_{\mu\nu}, b_\rho^\prime \}_P$ etc.
  
Using the Hamiltonian $H_T$, Eq. (\ref{Second1}) provides us with a secondary constraint:
\begin{eqnarray}
\Psi_2 = \partial_i \pi_S^i + ( 6 \xi + \epsilon ) 
( \pi_B - \tilde g^{0 \mu} \phi^2 S_\mu ) \approx 0,
\label{Second2}  
\end{eqnarray}
which is just the same as the $(0 \mu)$-components of the field equation for $S_\mu$ in (\ref{q-field-eq}) 
and there are no more constraints since we can show that\footnote{A derivation of constraints is exhibited in Appendix A.} 
\begin{eqnarray}
\dot \Psi_2  = \{ \Psi_2, H_T \}_P = 0.
\label{Tert}  
\end{eqnarray}

The Poisson bracket between the constraints is evaluated to be:
\begin{eqnarray}
\{ \Psi_1, \Psi_2^\prime \}_P = ( 6 \xi + \epsilon ) \tilde g^{00} \phi^2 \delta^3 
= ( 6 \xi + \epsilon ) \frac{1}{\tilde f} \phi^2 \delta^3,
\label{PB-12C}  
\end{eqnarray}
which implies that the constraints are the second-class constraint so that they can be treated 
by means of the Dirac bracket defined as
\begin{eqnarray}
\{ A, B^\prime \}_D \equiv \{ A, B^\prime \}_P - \{ A, \Psi_a^{\prime\prime} \}_P C_{ab}^{-1}
\{ \Psi_b^{\prime\prime}, B^\prime \}_P,  
\label{DB}  
\end{eqnarray}
where $\Psi_a ( a = 1, 2 )$ are the second-class constraints and $C_{ab}^{-1}$ is
the inverse matrix of $C_{ab} = \{ \Psi_a, \Psi_b^\prime \}_P$.
Concretely, the matrix elements, $C_{ab}^{-1}$, are given by
\begin{eqnarray}
C_{12}^{-1} = - C_{21}^{-1} = - \frac{1}{6 \xi + \epsilon} \tilde f \phi^{-2} \delta^3, \qquad
C_{11}^{-1} = C_{22}^{-1} = 0.
\label{C-1}  
\end{eqnarray}

As is well known, the canonical quantization can be carried out by replacing $i \{ A, B^\prime \}_D$
with the equal-time commutation relation $[ A, B^\prime ]$. After some calculations, we can 
write down several important ETCRs, which are needed for later calculations:
\begin{eqnarray} 
&{}& [ \dot g_{\rho\sigma}, g_{\mu\nu}^\prime ] = - \frac{2}{\xi} i \tilde f \phi^{-2} 
[ g_{\rho\sigma} g_{\mu\nu} - g_{\rho\mu} g_{\sigma\nu} - g_{\rho\nu} g_{\sigma\mu} 
+ \sqrt{-g} \tilde f ( \delta_\rho^0 \delta_\mu^0 g_{\sigma\nu} 
\nonumber\\
&{}& + \delta_\rho^0 \delta_\nu^0 g_{\sigma\mu} + \delta_\sigma^0 \delta_\mu^0 g_{\rho\nu}
+ \delta_\sigma^0 \delta_\nu^0 g_{\rho\mu} ) ] \delta^3, 
\nonumber\\
&{}&
[ \dot g_{\rho\sigma}, \phi^\prime ] = 0, \qquad 
[ \dot g_{\rho\sigma}, B^\prime ] = 2 i \tilde f \phi^{-2} g_{\rho\sigma} \delta^3,  
\nonumber\\
&{}& 
[ \phi, b_\rho^\prime ] =  [ B, b_\rho^\prime ] = [ B, \dot B^\prime ] = [ \dot \phi, \phi^\prime ] 
= [ \dot \phi, S_\mu^\prime ] = 0,   \qquad
[ \dot \phi, B^\prime ] = - i \tilde f \phi^{-1} \delta^3,  
\nonumber\\
&{}& 
[ S_0, S_k^\prime ] = - \frac{1}{6 \xi + \epsilon} i \tilde f \phi^{-2} \partial_k \delta^3, \qquad
[ S_k, S_l^\prime ] = 0, 
\nonumber\\
&{}& [ S_0, \dot S_k^\prime ] = - i \tilde f g_{0k} \delta^3,  \qquad
[ S_k, \dot S_l^\prime ] = - i \tilde f g_{kl} \delta^3 + \frac{1}{ 6 \xi + \epsilon } i
\partial_k ( \tilde f \phi^{-2} \partial_l \delta^3 ),
\nonumber\\
&{}& [ S_0, b_\mu^\prime ] = - i \tilde f \phi^{-2} S_\mu \delta^3, \qquad
[ S_0, B^\prime ] = - i \tilde f \phi^{-2} \delta^3,  \qquad
[ S_k, b_\rho^\prime ] = [ S_k, B^\prime ] = 0, 
\nonumber\\
&{}&
[ \dot S_0, B^\prime ] = 2 i \tilde f \tilde g^{0i} \partial_i ( \tilde f \phi^{-2} \delta^3 ), \qquad
[ \dot S_k, B^\prime ] = - i \partial_k ( \tilde f \phi^{-2} \delta^3 ),
\nonumber\\
&{}&
\{ \dot{\bar c}_\lambda, c^{\sigma \prime} \} = - \{ \dot c^\sigma, \bar c_\lambda^\prime \}
= - \tilde f \phi^{-2} \delta_\lambda^\sigma \delta^3,  \qquad
\{ \dot{\bar c}, c^\prime \} = - \{ \dot c, \bar c^\prime \}
= - \tilde f \phi^{-2} \delta^3, 
\nonumber\\
&{}& [ g_{\mu\nu}, b_\rho^\prime ] = - i \tilde f \phi^{-2} ( \delta_\mu^0 g_{\rho\nu} 
+ \delta_\nu^0 g_{\rho\mu} ) \delta^3,    \quad
[ \tilde g^{\mu\nu}, b_\rho^\prime ] = i \tilde f \phi^{-2} ( \tilde g^{\mu0} \delta_\rho^\nu 
+ \tilde g^{\nu0} \delta_\rho^\mu - \tilde g^{\mu\nu} \delta_\rho^0 ) \delta^3,
\nonumber\\
&{}& [ g_{\mu\nu}, \dot b_\rho^\prime ] = i \{ [ \tilde f \phi^{-2} \partial_\rho g_{\mu\nu} 
- \partial_0 ( \tilde f \phi^{-2} ) ( \delta_\mu^0 g_{\rho\nu} + \delta_\nu^0 g_{\rho\mu} ) ] \delta^3
+ [ ( \delta_\mu^k - 2 \delta_\mu^0 \tilde f \tilde g^{0k} ) g_{\rho\nu} 
\nonumber\\
&{}& + ( \mu \leftrightarrow \nu ) ] \partial_k (  \tilde f \phi^{-2} \delta^3 ) \},
\nonumber\\
&{}& [ \dot \phi, b_\rho^\prime ] = - i \tilde f \phi^{-2} \partial_\rho \phi \delta^3,    \qquad
[ \dot B, b_\rho^\prime ] = - i \tilde f \phi^{-2} \partial_\rho B \delta^3,
\nonumber\\
&{}& [ \dot S_0, b_\mu^\prime ] = - i \tilde f^2 \phi^{-2} [ \tilde g^{0\nu} ( \partial_\mu S_\nu
+ \partial_\nu S_\mu )  - \tilde g^{0i} H_{\mu i} ] \delta^3
+ 2 i \tilde f \tilde g^{0i} \partial_i ( \tilde f \phi^{-2} S_\mu \delta^3 ),
\nonumber\\
&{}& [ \dot S_k, b_\mu^\prime ] = i \tilde f \phi^{-2} H_{k \mu} \delta^3
- i \partial_k ( \tilde f \phi^{-2} S_\mu \delta^3 ),
\nonumber\\
&{}& [ b_\mu, b_\nu^\prime ] = 0,   \qquad
[ b_\mu, \dot b_\nu^\prime ] = i \tilde f \phi^{-2} ( \partial_\mu b_\nu + \partial_\nu b_\mu ) \delta^3,
\nonumber\\
&{}& [ b_\rho, c^{\sigma\prime} ] = [ b_\rho, \bar c_\lambda^\prime ] = [ b_\rho, c^\prime ] 
= [ b_\rho, \bar c^\prime ] = 0, 
\nonumber\\
&{}& [ \dot{\bar c}_\lambda, b_\rho^\prime ] = - i \tilde f \phi^{-2} \partial_\rho \bar c_\lambda \delta^3,  \qquad
[ \dot c^\sigma, b_\rho^\prime ] = - i \tilde f \phi^{-2} \partial_\rho c^\sigma \delta^3,
\nonumber\\
&{}& [ \dot{\bar c}, b_\rho^\prime ] = - i \tilde f \phi^{-2} \partial_\rho \bar c \delta^3, \qquad
[ \dot c, b_\rho^\prime ] = - i \tilde f \phi^{-2} \partial_\rho c \delta^3.
\label{ETCRs}  
\end{eqnarray}
These ETCRs can be obtained from the explicit calculations and/or the BRST transformations.
For instance, we will present a derivation of $[ B, \dot B^\prime ] = 0$ by the both methods.
First, let us focus on the explicit calculation via the Dirac bracket:
\begin{eqnarray}
\{ B, \dot B^\prime \}_D = \{ B, \dot B^\prime \}_P - \{ B, \Psi_2^{\prime\prime} \}_P C_{21}^{-1}
\{ \Psi_1^{\prime\prime}, \dot B^\prime \}_P.
\label{B-B'}  
\end{eqnarray}
Since we can easily evaluate each Poisson bracket whose result reads:
\begin{eqnarray}
\{ B, \dot B^\prime \}_P &=& \{ B, ( 6 \xi + \epsilon ) \tilde f \phi^{-2} \pi_B^\prime \}_P
= ( 6 \xi + \epsilon ) \tilde f \phi^{-2} \delta^3,
\nonumber\\
\{ B, \Psi_2^\prime \}_P  &=& \{ B, ( 6 \xi + \epsilon ) \pi_B^\prime \}_P
= ( 6 \xi + \epsilon ) \delta^3,
\nonumber\\
\{ \Psi_1, \dot B^\prime \}_P &=& \{ \pi_S^0, - ( 6 \xi + \epsilon ) \tilde f \tilde g^{0\mu} S_\mu^\prime \}
= ( 6 \xi + \epsilon ) \delta^3,
\label{B-B'2}  
\end{eqnarray}
the Dirac bracket becomes:
\begin{eqnarray}
\{ B, \dot B^\prime \}_D &=& ( 6 \xi + \epsilon ) \tilde f \phi^{-2} \delta^3
- ( 6 \xi + \epsilon ) C_{21}^{-1} ( 6 \xi + \epsilon ) \delta^3
\nonumber\\
&=& 0.
\label{B-B'3}  
\end{eqnarray}

Second, we will present a derivation by means of the BRST transformation which is more general and elegant
than the above explicit calculation. The ETCR, $[ B, \pi_c^\prime ] = 0$, leads to $[ B, \dot{\bar{c}}^\prime ] = 0$.
Taking the Weyl BRST transformation of this ETCR yields the equation:
\begin{eqnarray}
\{ [ i \bar Q_B, B ], \dot{\bar{c}}^\prime \} + [ B, \{ i \bar Q_B, \dot{\bar{c}}^\prime \} ] = 0.
\label{W-B-B'}  
\end{eqnarray}
Then, the Weyl BRST transformation (\ref{Weyl-BRST}) immediately leads to $[ B, \dot B^\prime ] = 0$.

\section{Unitarity analysis}

As in the conventional BRST formalism, the physical state $| \rm{phys} \rangle$ is defined by
imposing two subsidiary conditions \cite{Kugo-Ojima}:
\begin{eqnarray}
Q_B | \rm{phys} \rangle = \bar Q_B | \rm{phys} \rangle = 0.
\label{Phys-state}  
\end{eqnarray}
It is then well known that the physical S-matrix is unitary under the assumption that all 
BRST singlet states have positive norm. In this section, we would like to prove the unitarity of the
physical S-matrix in the present theory. From the classical analysis we know that the gauge field becomes
massive via the Higgs mechanism. Thus, we wish to understand how the Higgs mechanism is described 
in terms of the BRST formalism.

In analysing the unitarity, it is enough to take account of asymptotic fields of all the fundamental
fields and the free part of the Lagrangian. Let us first assume the asymptotic fields as
\begin{eqnarray}
g_{\mu\nu} &=& \eta_{\mu\nu} + \varphi_{\mu\nu},  \qquad
\phi = \phi_0 + \tilde \phi, \qquad
S_\mu = s_\mu,  \qquad
b_\mu = \beta_\mu, \qquad
B = \beta,
\nonumber\\
c^\mu &=& \gamma^\mu, \qquad
\bar c_\mu = \bar \gamma_\mu, \qquad
c = \gamma, \qquad
\bar c = \bar \gamma,
\label{Asmp-exp}  
\end{eqnarray}
where $\eta_{\mu\nu} ( = \eta^{\mu\nu} )$ is the flat Minkowski metric with the mostly positive signature 
and $\phi_0$ is a non-zero constant. In this section, the Minkowski metric is used to lower or raise the 
Lorentz indices. Using these asymptotic fields, the free part of the Lagrangian reads:
\begin{eqnarray}
{\cal L}_q &=& \frac{1}{2} \xi \phi_0^2 \left( \frac{1}{4} \varphi_{\mu\nu} \Box \varphi^{\mu\nu} 
- \frac{1}{4} \varphi \Box \varphi - \frac{1}{2} \varphi^{\mu\nu} \partial_\mu \partial_\rho \varphi_\nu{}^\rho
+ \frac{1}{2} \varphi^{\mu\nu} \partial_\mu \partial_\nu \varphi \right)
\nonumber\\
&+& \xi \phi_0 \tilde \phi \left( - \Box \varphi + \partial_\mu \partial_\nu \varphi^{\mu\nu} \right)
+ ( 6 \xi + \epsilon ) \left( \phi_0 S_\mu \partial^\mu \tilde \phi - \frac{1}{2} \phi_0^2 S_\mu S^\mu \right)
\nonumber\\
&-& \frac{1}{4} h_{\mu\nu}^2 
- \frac{1}{2} \epsilon \partial_\mu \tilde \phi \partial^\mu \tilde \phi
- \left( 2 \eta^{\mu\nu} \phi_0 \tilde \phi - \phi_0^2 \varphi^{\mu\nu} + \frac{1}{2} 
\phi_0^2 \eta^{\mu\nu} \varphi \right) \partial_\mu \beta_\nu
\nonumber\\
&-& i \phi_0^2 \partial_\mu \bar \gamma_\rho \partial^\mu \gamma^\rho
+ \phi_0 \partial_\mu \beta \partial^\mu \tilde \phi - i \phi_0^2 \partial_\mu \bar \gamma \partial^\mu \gamma,
\label{Free-Lag}  
\end{eqnarray}
where $\Box \equiv \eta^{\mu\nu} \partial_\mu \partial_\nu, \, \varphi \equiv \eta^{\mu\nu} \varphi_{\mu\nu}$
and $h_{\mu\nu} \equiv \partial_\mu s_\nu - \partial_\nu s_\mu$.
Based on this Lagrangian, it is easy to derive the linearized field equations: 
\begin{eqnarray}
&{}& \frac{1}{2} \xi \phi_0^2 \biggl( \frac{1}{2} \Box \varphi_{\mu\nu} - \frac{1}{2} \eta_{\mu\nu} \Box \varphi 
- \partial_\rho \partial_{(\mu} \varphi_{\nu)}{}^\rho + \frac{1}{2} \partial_\mu \partial_\nu \varphi
+ \frac{1}{2} \eta_{\mu\nu} \partial_\rho \partial_\sigma \varphi^{\rho\sigma} \biggr)
\nonumber\\
&{}& + \xi \phi_0 \left( - \eta_{\mu\nu} \Box + \partial_\mu \partial_\nu \right) \tilde \phi 
+\phi_0^2 \partial_{(\mu} \beta_{\nu)} - \frac{1}{2} \phi_0^2 \eta_{\mu\nu} \partial_\rho \beta^\rho = 0.
\label{Linear-Eq1}
\\
&{}& \epsilon \Box \tilde \phi + \xi \phi_0 ( - \Box \varphi + \partial_\mu \partial_\nu \varphi^{\mu\nu} )
- ( 6 \xi + \epsilon ) \phi_0 \partial_\mu s^\mu - 2 \phi_0 \partial_\rho \beta^\rho 
\nonumber\\
&{}& - \phi_0 \Box \beta = 0.
\label{Linear-Eq2}
\\
&{}& \partial^\nu h_{\mu\nu} + ( 6 \xi + \epsilon ) \phi_0^2 \left( s_\mu - \frac{1}{\phi_0} 
\partial_\mu \tilde \phi \right) = 0.
\label{Linear-Eq3}
\\
&{}& \partial_\mu \tilde \phi - \frac{1}{2} \phi_0 \left( \partial^\nu \varphi_{\mu\nu} 
- \frac{1}{2} \partial_\mu \varphi \right) = 0.
\label{Linear-Eq4}
\\
&{}& \Box \tilde \phi = \Box \gamma^\mu =  \Box \bar \gamma_\mu = \Box \gamma 
= \Box \bar \gamma = 0. 
\label{Linear-Eq5}  
\end{eqnarray}
Here we have introduced the symmetrization notation $A_{(\mu} B_{\nu)} \equiv \frac{1}{2} ( A_\mu
B_\nu + A_\nu B_\mu )$. 

Now, operating $\partial^\mu$ on Eq. (\ref{Linear-Eq4}) and using Eq. (\ref{Linear-Eq5}), we obtain:
\begin{eqnarray}
\partial_\mu \partial_\nu \varphi^{\mu\nu} - \frac{1}{2} \Box \varphi = 0.
\label{Linear-Eq6}  
\end{eqnarray}
Next, taking the trace of Eq. (\ref{Linear-Eq1}) with the help of Eqs. (\ref{Linear-Eq5}) and 
(\ref{Linear-Eq6}) leads to:
\begin{eqnarray}
\Box \varphi + \frac{4}{\xi} \partial_\rho \beta^\rho = 0.
\label{Linear-Eq7}  
\end{eqnarray}
Moreover, operating $\partial^\mu$ on Eq. (\ref{Linear-Eq3}), and using the identity $\partial^\mu \partial^\nu 
h_{\mu\nu} = 0$ and Eq. (\ref{Linear-Eq5}) yields the Lorenz condition:
\begin{eqnarray}
\partial_\mu s^\mu = 0.
\label{Linear-Lorenz}  
\end{eqnarray}
As can been seen in Eq. (\ref{Linear-Eq3}), it is more convenient to introduce $\hat s_\mu$
defined as
\begin{eqnarray}
\hat s_\mu = s_\mu - \frac{1}{\phi_0} \partial_\mu \tilde \phi,
\label{hat-s}  
\end{eqnarray}
which also obeys the Lorenz condition owing to Eqs. (\ref{Linear-Eq5}) and (\ref{Linear-Lorenz}):
\begin{eqnarray}
\partial_\mu \hat s^\mu = 0.
\label{Linear-Lorenz2}  
\end{eqnarray}
With the new gauge field $\hat s_\mu$ and the corresponding field strength $\hat h_{\mu\nu} \equiv
\partial_\mu \hat s_\nu - \partial_\nu \hat s_\mu$, the ``Maxwell equation'' (\ref{Linear-Eq3}) can be cast 
to the form:
\begin{eqnarray}
\partial^\nu \hat h_{\mu\nu} + ( 6 \xi + \epsilon ) \phi_0^2 \hat s_\mu = 0,
\label{Mass-Max}  
\end{eqnarray}
which clearly shows that the Weyl gauge field absorbs the Nambu-Goldstone boson $\tilde \phi$
associated with spontaneous symmetry breakdown of the Weyl gauge symmetry, thereby becoming
massive with the mass squared $( 6 \xi + \epsilon ) \phi_0^2$ (Here we assume $6 \xi + \epsilon > 0$,
which is consistent with the positive Newton constant $\xi > 0$).  To put it differently,
after spontaneous symmetry breakdown of the Weyl gauge symmetry, the Weyl gauge field $\hat s_\mu$ satisfies 
not only the Lorenz condition (\ref{Linear-Lorenz2}) but also the massive Klein-Gordon equation:
\begin{eqnarray}
( \Box - m^2 ) \hat s_\mu = 0,
\label{Mass-KG}  
\end{eqnarray}
where $m^2$ is defined by
\begin{eqnarray}
m^2 \equiv ( 6 \xi + \epsilon ) \phi_0^2.
\label{Mass-squared}  
\end{eqnarray}

Furthermore, with the help of Eqs. (\ref{Linear-Eq5}), (\ref{Linear-Eq6}), (\ref{Linear-Eq7}) and (\ref{Linear-Lorenz}),
Eq. (\ref{Linear-Eq2}) can be rewritten as
\begin{eqnarray}
\Box \beta = 0.
\label{Linear-Eq8}  
\end{eqnarray}
Moreover, acting $\partial^\mu$ on Eq. (\ref{Linear-Eq1}) yields:
\begin{eqnarray}
\Box \beta_\mu = 0.
\label{Linear-Eq9}  
\end{eqnarray}
Finally, using various equations obtained thus far, the ``Einstein equation'' (\ref{Linear-Eq1}) is reduced 
to the form:
\begin{eqnarray}
\Box \varphi_{\mu\nu} + \frac{4}{\xi} \partial_{(\mu} \beta_{\nu)} = 0,
\label{Linear-Eq10}  
\end{eqnarray}
which means that the field $\varphi_{\mu\nu}$ is not a simple pole field but a dipole field:
\begin{eqnarray}
\Box^2 \varphi_{\mu\nu} = 0.
\label{Linear-Eq11}  
\end{eqnarray}
On the other hand, in addition to Eq. (\ref{Mass-KG}), the other fields are all simple pole fields:
\begin{eqnarray}
\Box \tilde \phi = \Box \beta_\mu = \Box \beta =\Box \gamma^\mu =  \Box \bar \gamma_\mu 
= \Box \gamma = \Box \bar \gamma = 0. 
\label{Linear-Eq12}  
\end{eqnarray}
Note that Eq. (\ref{Linear-Eq12}) corresponds to Eq. (\ref{X-M-eq}) in a curved space-time.

Following the standard technique, we can calculate the four-dimensional (anti-)commutation 
relations (4D CRs) between asymptotic fields. The point is that the simple pole fields, for instance, 
the Nakanishi-Lautrup field $\beta_\mu (x)$ can be expressed in terms of the invariant delta
function $D(x)$ as
\begin{eqnarray}
\beta_\mu (x) = - \int d^3 z D(x-z) \overleftrightarrow{\partial}_0^z \beta_\mu (z),
\label{D-beta}  
\end{eqnarray}
whereas the dipole field $\varphi_{\mu\nu}(x)$ takes the form:
\begin{eqnarray}
&{}& \varphi_{\mu\nu} (x) = - \int d^3 z \left[ D(x-z) \overleftrightarrow{\partial}_0^z \varphi_{\mu\nu} (z)
+ E(x-z) \overleftrightarrow{\partial}_0^z \Box \varphi_{\mu\nu} (z) \right]
\nonumber\\
&{}& = - \int d^3 z \left[ D(x-z) \overleftrightarrow{\partial}_0^z \varphi_{\mu\nu} (z)
- \frac{4}{\xi} E(x-z) \overleftrightarrow{\partial}_0^z \partial_{(\mu} \beta_{\nu)} (z) \right],
\label{E-varphi}  
\end{eqnarray}
where in the last equality we have used Eq. (\ref{Linear-Eq10}).
Here the invariant delta function $D(x)$ for massless simple pole fields and its properties
are described as
\begin{eqnarray}
&{}& D(x) = - \frac{i}{(2 \pi)^3} \int d^4 k \, \epsilon (k_0) \delta (k^2) e^{i k x}, \qquad
\Box D(x) = 0,
\nonumber\\
&{}& D(-x) = - D(x), \qquad D(0, \vec{x}) = 0, \qquad 
\partial_0 D(0, \vec{x}) = \delta^3 (x), 
\label{D-function}  
\end{eqnarray}
where $\epsilon (k_0) \equiv \frac{k_0}{|k_0|}$. Similarly, the invariant delta function $E(x)$ 
for massless dipole fields and its properties are given by
\begin{eqnarray}
&{}& E(x) = - \frac{i}{(2 \pi)^3} \int d^4 k \, \epsilon (k_0) \delta^\prime (k^2) e^{i k x}, \qquad  
\Box E(x) = D(x),
\nonumber\\
&{}& E(-x) = - E(x), \qquad 
E(0, \vec{x}) = \partial_0 E(0, \vec{x}) = \partial_0^2 E(0, \vec{x}) = 0, 
\nonumber\\ 
&{}& \partial_0^3 E(0, \vec{x}) = - \delta^3 (x),
\label{E-function}  
\end{eqnarray}
where $\delta^\prime (k^2) \equiv \frac{d \delta (k^2)}{d k^2}$.

On the other hand, the Weyl gauge field $\hat s(x)$ obeys the massive Klein-Gordon equation
(\ref{Mass-KG}), so it needs to be described in terms of the invariant delta function 
$\Delta(x; m^2)$ for massive simple pole fields as
\begin{eqnarray}
\hat s_\mu (x) = - \int d^3 z \Delta (x-z; m^2) \overleftrightarrow{\partial}_0^z \hat s_\mu (z),
\label{s-Delta}  
\end{eqnarray}
where $\Delta(x; m^2)$ is defined as
\begin{eqnarray}
&{}& \Delta(x; m^2) = - \frac{i}{(2 \pi)^3} \int d^4 k \, \epsilon (k_0) \delta (k^2 + m^2) e^{i k x}, \quad
(\Box - m^2) \Delta(x; m^2) = 0,
\nonumber\\
&{}& \Delta(-x; m^2) = - \Delta(x; m^2), \quad \Delta(0, \vec{x}; m^2) = 0, 
\nonumber\\
&{}& \partial_0 \Delta(0, \vec{x}; m^2) = \delta^3 (x),  \qquad
\Delta(x; 0) = D(x). 
\label{Delta-function}  
\end{eqnarray}
It is easy to show that the RHS of Eqs. (\ref{D-beta}), (\ref{E-varphi}) and (\ref{s-Delta}) is independent of
$z^0$. Thus, for instance, when we evaluate the four-dimensional commutation relation
$[ \varphi_{\mu\nu} (x), \varphi_{\sigma\tau} (y) ]$, we can put $z^0 = y^0$ and use the 
three-dimensional commutation relations among asymptotic fields. After some manipulation, we find that
the 4D CRs are given by
\begin{eqnarray}
&{}& [ \varphi_{\mu\nu} (x), \varphi_{\sigma\tau} (y) ] = - \frac{2}{\xi} i \phi_0^{-2} [ ( \eta_{\mu\nu} \eta_{\sigma\tau}
- \eta_{\mu\sigma} \eta_{\nu\tau} - \eta_{\mu\tau} \eta_{\nu\sigma} ) D(x-y) 
\nonumber\\
&{}& + ( \eta_{\mu\sigma} \partial_\nu \partial_\tau + \eta_{\nu\sigma} \partial_\mu \partial_\tau +
\eta_{\mu\tau} \partial_\nu \partial_\sigma + \eta_{\nu\tau} \partial_\mu \partial_\sigma ) E(x-y) ],
\label{4D-CR1}
\\
&{}& [ \varphi_{\mu\nu} (x), \beta_\rho (y) ] = - i \phi_0^{-2} ( \eta_{\mu\rho} \partial_\nu
+ \eta_{\nu\rho} \partial_\mu ) D(x-y). 
\label{4D-CR2}
\\
&{}& [ \varphi_{\mu\nu} (x), \beta (y) ] = 2 i \phi_0^{-1} \eta_{\mu\nu} D(x-y). 
\label{4D-CR3}
\\
&{}& [ \tilde \phi (x), \beta (y) ] = - i \phi_0^{-1} D(x-y). 
\label{4D-CR4}
\\
&{}& [ \hat s_\mu (x), \hat s_\nu (y) ] = i \left( \eta_{\mu\nu} - \frac{1}{m^2} \partial_\mu \partial_\nu \right) 
\Delta (x-y; m^2). 
\label{4D-CRs1}
\\
&{}& \{ \gamma^\sigma (x), \bar \gamma_\tau (y) \} = \phi_0^{-2} \delta_\tau^\sigma D(x-y). 
\label{4D-CR5}
\\
&{}& \{ \gamma (x), \bar \gamma (y) \} = \phi_0^{-2} D(x-y). 
\label{4D-CR6}  
\end{eqnarray}
The other 4D CRs vanish identically.

Now we would like to discuss the issue of the unitarity of the physical S-matrix. To do that, it is
convenient to perform the Fourier transformation of Eqs. (\ref{4D-CR1})-(\ref{4D-CR6}).
However, for the dipole field we cannot use the three-dimensional Fourier expansion to define 
the creation and annihilation operators. We therefore make use of the four-dimensional 
Fourier expansion \cite{N-O-text}:\footnote{For simplicity, the Fourier transform of a field is denoted 
by the same field except for the argument $p$ instead of $x$.}
\begin{eqnarray}
\varphi_{\mu\nu} (x) = \frac{1}{(2 \pi)^{\frac{3}{2}}} \int d^4 p \, \theta (p_0) [ \varphi_{\mu\nu} (p) e^{i p x}
+ \varphi_{\mu\nu}^\dagger (p) e^{- i p x} ],
\label{FT-varphi}  
\end{eqnarray}
where $\theta (p_0)$ is the step function. For any simple pole fields, we adopt the same Fourier expansion,
for instance, 
\begin{eqnarray}
\beta_\mu (x) = \frac{1}{(2 \pi)^{\frac{3}{2}}} \int d^4 p \, \theta (p_0) [ \beta_\mu (p) e^{i p x}
+ \beta_\mu^\dagger (p) e^{- i p x} ].
\label{FT-beta}  
\end{eqnarray}
Thus, using Eqs. (\ref{D-beta}), (\ref{E-varphi}), (\ref{FT-varphi}) and (\ref{FT-beta}), for instance, 
the Fourier transforms of, e.g., $\varphi_{\mu\nu} (x)$ and $\beta_\mu (x)$ take the following expression:  
\begin{eqnarray}
\varphi_{\mu\nu} (p) &=& \frac{i}{(2 \pi)^{\frac{3}{2}}} \theta( p_0 ) \int d^3 z \, e^{-i p z} 
\overleftrightarrow{\partial}_0^z [ \delta(p^2) \varphi_{\mu\nu} (z) + \delta^\prime (p^2) 
\Box \varphi_{\mu\nu} (z) ],
\nonumber\\
\beta_\mu (p) &=& \frac{i}{(2 \pi)^{\frac{3}{2}}} \theta( p_0 ) \delta(p^2) \int d^3 z \, e^{-i p z} 
\overleftrightarrow{\partial}_0^z \beta_\mu (z).
\label{FT-fields}  
\end{eqnarray}

Incidentally, for a generic simple pole field $\Phi$ with a mass $m$, the three-dimensional Fourier expansion 
is defined as
\begin{eqnarray}
\Phi (x) = \frac{1}{(2 \pi)^{\frac{3}{2}}} \int d^3 p \, \frac{1}{\sqrt{2 \omega_p}}  
[ \Phi (\vec{p}) e^{i p x} + \Phi^\dagger (\vec{p}) e^{ - i p x } ],
\label{3D-FT}  
\end{eqnarray}
with being $\omega_p = \sqrt{ \vec{p}^2 + m^2}$, whereas the four-dimensional Fourier expansion reads:
\begin{eqnarray}
\Phi (x) = \frac{1}{(2 \pi)^{\frac{3}{2}}} \int d^4 p \, \theta (p_0) [ \Phi (p) e^{i p x}
+ \Phi^\dagger (p) (p) e^{- i p x} ].
\label{4D-FT}  
\end{eqnarray}
Thus, the annihilation operator $\Phi (p)$ in the four-dimensional Fourier expansion has connection with 
the annihilation operator $\Phi (\vec{p})$ in the three-dimensional Fourier expansion via
\begin{eqnarray}
\Phi (p) = \theta (p_0) \delta (p^2 + m^2) \sqrt{2 \omega_p} \Phi (\vec{p}).
\label{3D-4D}  
\end{eqnarray}
Based on these Fourier expansions, we can calculate the Fourier transform of Eqs. (\ref{4D-CR1})-(\ref{4D-CR6}):
\begin{eqnarray}
&{}& [ \varphi_{\mu\nu} (p), \varphi_{\sigma\tau}^\dagger (q) ] = - \frac{2}{\xi} \phi_0^{-2} \theta (p_0) \delta^4 (p-q)
[ \delta(p^2) ( \eta_{\mu\nu} \eta_{\sigma\tau}- \eta_{\mu\sigma} \eta_{\nu\tau} - \eta_{\mu\tau} \eta_{\nu\sigma} ) 
\nonumber\\
&{}& - 3  \delta^\prime (p^2) ( \eta_{\mu\sigma} p_\nu p_\tau + \eta_{\nu\sigma} p_\mu p_\tau +
\eta_{\mu\tau} p_\nu p_\sigma + \eta_{\nu\tau} p_\mu p_\sigma ) ].
\label{FT-4D-CR1}
\\
&{}& [ \varphi_{\mu\nu} (p), \beta_\rho^\dagger (q) ] = - i \phi_0^{-2} ( \eta_{\mu\rho} p_\nu + \eta_{\nu\rho} p_\mu ) 
\theta (p_0) \delta(p^2) \delta^4 (p-q). 
\label{FT-4D-CR2}
\\
&{}& [ \varphi_{\mu\nu} (p), \beta^\dagger (q) ] = 2 \phi_0^{-1} \eta_{\mu\nu} \theta (p_0) \delta(p^2) \delta^4 (p-q). 
\label{FT-4D-CR3}
\\
&{}& [ \tilde \phi (p), \beta^\dagger (q) ] = - \phi_0^{-1} \theta (p_0) \delta(p^2) \delta^4 (p-q). 
\label{FT-4D-CR4}
\\
&{}& [ \hat s_\mu (p), \hat s_\nu^\dagger (q) ] = + \left( \eta_{\mu\nu} - \frac{1}{m^2} p_\mu p_\nu \right) 
\theta(p_0) \delta(p^2 + m^2) \delta^4 (p-q). 
\label{FT-4D-CRs}
\\
&{}& \{ \gamma^\sigma (p), \bar \gamma^\dagger_\tau (q) \} = - i \phi_0^{-2} \delta_\tau^\sigma \theta (p_0) 
\delta(p^2) \delta^4 (p-q). 
\label{FT-4D-CR5}
\\
&{}& \{ \gamma (p), \bar \gamma^\dagger (q) \} = - i \phi_0^{-2} \theta (p_0) \delta(p^2) \delta^4 (p-q). 
\label{FT-4D-CR6}  
\end{eqnarray}

Next, let us turn our attention to the linearized field equations. After Fourier transformation, 
Eq. (\ref{Linear-Eq4}) takes the form:
\begin{eqnarray}
p^\nu \varphi_{\mu\nu} - \frac{1}{2} p_\mu \varphi = 2 \phi_0^{-1} p_\mu \tilde \phi.
\label{FT-Linear-Eq3}
\end{eqnarray}
If we fix the degree of freedom associated with $\tilde \phi$, which will be discussed later,
this equation gives us four independent relations on ten components of $\varphi_{\mu\nu} (p)$,
thereby reducing the independent components of $\varphi_{\mu\nu} (p)$ to be six. To deal with
six independent components of $\varphi_{\mu\nu} (p)$, it is convenient to take a specific Lorentz 
frame such that $p_1 = p_2 = 0$ and $p_3 > 0$, and choose the six components as follows:
\begin{eqnarray}
&{}& \varphi_1 (p) = \frac{1}{2} [ \varphi_{11} (p) - \varphi_{22} (p) ],  \qquad
\varphi_2 (p) = \varphi_{12} (p),  \qquad
\omega_0 (p) = - \frac{1}{2 p_0} \varphi_{00} (p),   
\nonumber\\
&{}& \omega_I (p) = - \frac{1}{p_0} \varphi_{0I} (p),  \qquad
\omega_3 (p) = - \frac{1}{2 p_3} \varphi_{33} (p), 
\label{Lorentz}  
\end{eqnarray}
where the index $I$ takes the transverse components $I = 1, 2$. 

In this respect, it is worthwhile to consider the GCT BRST transformation for these components.
First, let us write down the GCT BRST transformation for the Fourier expansion of the asymptotic fields, 
which reads:
\begin{eqnarray}
&{}& \delta_B \varphi_{\mu\nu} (p) = - i [ p_\mu \gamma_\nu (p) + p_\nu \gamma_\mu (p) ], \quad
\delta_B \gamma^\mu (p) = 0, \quad 
\delta_B \bar \gamma_\mu (p) = i \beta_\mu (p),
\nonumber\\
&{}& \delta_B \tilde \phi (p) = \delta_B \beta_\mu (p) = \delta_B \beta (p) 
= \delta_B \gamma (p) = \delta_B \bar \gamma (p) = 0. 
\label{Q_B-FT}  
\end{eqnarray}
Using this BRST transformation, the GCT BRST transformation for the components
in (\ref{Lorentz}) takes the form:
\begin{eqnarray}
&{}& \delta_B \varphi_I (p) = 0, \qquad
\delta_B \omega_\mu (p) = i \gamma_\mu (p),
\nonumber\\
&{}& \delta_B \bar \gamma_\mu (p) = i \beta_\mu (p), \qquad 
\delta_B \gamma_\mu (p) = \delta_B \beta_\mu (p) = 0,
\label{Q_B-Comp}  
\end{eqnarray}
where $p_1 = p_2 = 0$ was used. This BRST transformation implies that $\varphi_I (p)$
could be the physical observable while a set of fields, $\{ \omega_\mu (p), \beta_\mu (p), 
\gamma_\mu (p), \bar \gamma_\mu (p) \}$ might belong to the BRST quartet and thus are dropped
from the physical state by the Kugo-Ojima subsidiary condition, $Q_B | \rm{phys} \rangle = 0$ 
\cite{Kugo-Ojima}.\footnote{The situation is in fact a bit complicated since $\beta_\mu (p), \gamma_\mu (p)$ 
and $\bar \gamma_\mu (p)$ 
are simple pole fields obeying $p^2 \beta_\mu (p) = p^2 \gamma_\mu (p) = p^2 \bar \gamma_\mu (p) = 0$,
while $\varphi_{\mu\nu} (p)$ is a dipole field satisfying $( p^2 )^2 \varphi_{\mu\nu} (p) = 0$, 
so that a naive Kugo-Ojima's quartet mechanism does not work in a direct way. But this problem can be 
remedied by introducing an operator which takes out a simple pole from a dipole field. The detail can be 
shown in Ref. \cite{Oda-W}.} 

Next, let us move on to the other BRST transformation, which is the BRST transformation for the Weyl 
transformation. The Weyl BRST transformation for the asymptotic fields is of form:
\begin{eqnarray}
&{}& \bar \delta_B \varphi_{\mu\nu} = 2 c \eta_{\mu\nu}, \quad
\bar \delta_B \tilde \phi = - \phi_0 \gamma, \quad 
\bar \delta_B \gamma = 0, \quad 
\bar \delta_B \bar \gamma = i \beta,
\nonumber\\
&{}& \bar \delta_B \beta = \bar \delta_B \beta_\mu = \bar \delta_B \gamma_\mu 
= \bar \delta_B \bar \gamma_\mu = 0. 
\label{W-Q_B-Asym}  
\end{eqnarray}
The Weyl BRST transformation of $\varphi_I$ is vanishing:
\begin{eqnarray}
\bar \delta_B \varphi_I = 0, 
\label{W-Q_B-Obs}  
\end{eqnarray}
which means that together with $\delta_B \varphi_I = 0$, $\varphi_I$ is truely the physical observable.
The four-dimensional commutation relations among the fields $\{ \tilde \phi, \beta,
\gamma, \bar \gamma \}$ read:
\begin{eqnarray}
&{}& [ \tilde \phi (p), \tilde \phi^\dagger (q) ] = 0, 
\nonumber\\
&{}& [ \tilde \phi (p), \beta^\dagger (q) ] = - \phi_0^{-1} \theta (p_0) \delta (p^2) \delta^4 (p-q), 
\nonumber\\
&{}& \{ \gamma (p), \bar \gamma^\dagger (q) ] = - i \phi_0^{-2} \theta (p_0) \delta (p^2) \delta^4 (p-q).
\label{W-4D-CRs}  
\end{eqnarray}
As can be also seen in these 4D CRs, all the fields $\{ \varphi_I, \tilde \phi, \beta, \gamma, \bar \gamma \}$   
are massless simple pole fields. Via relation (\ref{3D-4D}) the three-dimensional commutation
relations $[ \Phi (\vec{p}), \Phi^\dagger (\vec{q}) \}$ with $\Phi (\vec{p}) \equiv \{ \varphi_I (\vec{p}), 
\tilde \phi (\vec{p}), \beta (\vec{p}), \gamma (\vec{p}), \bar \gamma (\vec{p}) \}$, are of form:  
\begin{eqnarray}
[ \Phi (\vec{p}), \Phi^\dagger (\vec{q}) \} &=&
\left(
\begin{array}{cc|cc|cc}
\frac{2}{\xi} \phi_0^{-2}  \delta_{IJ}              &     &    &    &         \\ 
\hline
    &        &               0   &    - \phi_0^{-1}     &            \\ 
    &        &               - \phi_0^{-1}    &  0       &              \\
\hline   
    &        &     &    &       &    -i \phi_0^{-2}  \\  
    &        &     &    &   +i \phi_0^{-2} &            \\
\end{array}
\right) 
\nonumber\\
&\times& \delta ( \vec{p} - \vec{q} ).
\label{W-3D-CRs}  
\end{eqnarray}
Thus, $\varphi_I$ is the physical observable while the set of fields, $\{ \tilde \phi, \beta, \gamma, \bar \gamma \}$ 
consists of the BRST quartet and is the unphysical mode by the Kugo-Ojima's subsidiary condition \cite{Kugo-Ojima}. 
Here it is worth mentioning that the Nambu-Goldstone boson $\tilde \phi$ associated with spontaneous symmetry
breaking of the Weyl gauge symmetry is an unphysical particle. In this context, let us recall that the Nambu-Goldstone theorem 
never tells us whether the Nambu-Goldstone boson is physical or unphysical. From our analysis at hand, we can conclude 
that the Nambu-Goldstone boson $\tilde \phi$ is the unphysical mode, which is absorbed into the longitudinal mode
of the Weyl gauge field $s_\mu(x)$, thereby the gauge field becoming massive.   

Finally, let us focus on the Weyl gauge field $\hat s_\mu$, which satisfies the Lorenz condition (\ref{Linear-Lorenz2}) 
and the massive Klein-Gordon equation (\ref{Mass-KG}). In a specific Lorentz frame: 
\begin{eqnarray}
p_\mu = ( m, 0, 0, 0),
\label{Lor-frame}  
\end{eqnarray}
the Lorenz condition (\ref{Linear-Lorenz2}) produces:
\begin{eqnarray}
\hat s_0 (p) = 0.
\label{Zero-s0}  
\end{eqnarray}
With the Lorentz frame (\ref{Lor-frame}), it turns out that the spacial components of $\hat s_\mu$ are
invariant under both GCT and Weyl BRST transformations:
\begin{eqnarray}
\delta_B \hat s_i (p) = \bar \delta_B \hat s_i (p) = 0.
\label{Zero-BRST}  
\end{eqnarray}
Moreover, using the relation (\ref{3D-4D}) and Eq. (\ref{FT-4D-CRs}), the commutation relation between 
the three-dimensional annihilation and creation operators reads:
\begin{eqnarray}
[ \hat s_i (\vec{p}), \hat s_j^\dagger (\vec{q}) ] = \delta_{ij} \delta^3 ( \vec{p} - \vec{q} ).
\label{3D-Rel}  
\end{eqnarray}
Together with the BRST invariance in Eq. (\ref{Zero-BRST}),  this equation clearly shows that the spacial
components $\hat s_i (x)$ are really genuine physical massive modes belonging to BRST singlets with
positive norm.

\section{Choral symmetry}

In the previous article \cite{Oda-W}, we have clarified the existence of a huge global symmetry 
called ``choral symmetry'', which is the $IOSp(10|10)$ symmetry, in Weyl invariant scalar-tensor gravity 
in Riemann geometry. We will show that the choral symmetry also exists in the theory at hand. 
The existence of the choral symmetry is expected from the fact that as shown in Section 4, a set of fields (including 
the space-time coordinates $x^\mu$) $X^M \equiv \{ x^\mu, b_\mu, \sigma, B, c^\mu, \bar c_\mu, c, \bar c \}$ 
obeys a very simple equation:
\begin{eqnarray}
g^{\mu\nu} \partial_\mu \partial_\nu X^M = 0.
\label{d'Alemb-eq}  
\end{eqnarray}
It is worthwhile to note that this equation holds if and only if we adopt the extended de Donder gauge 
condition (\ref{Ext-de-Donder}) for the GCT and the scalar gauge condition (\ref{Scalar-gauge}) for the Weyl gauge 
transformation. Furthermore, Eq. (\ref{d'Alemb-eq}) implies that there should be many conserved currents defined 
in Eq. (\ref{Cons-currents}) in the theory under consideration.  In this section, along the same line of argument 
as that in the previous article \cite{Oda-W, Oda-V}, we will explicitly prove that there is the choral symmetry
$IOSp(10|10)$ in Weyl conformal gravity in Weyl geometry.

Let us start with the Lagrangian (\ref{ST-q-Lag}), which can be cast to the form:
\begin{eqnarray}
{\cal L}_q &=& \sqrt{- g}  \biggl[  \frac{1}{2} \xi \phi^2 ( R - 6 \nabla_\mu S^\mu - 6 S_\mu S^\mu ) 
- \frac{1}{4} H_{\mu\nu} H^{\mu\nu} 
- \frac{1}{2} \epsilon g^{\mu\nu} ( - 2 \phi \partial_\mu \phi S_\nu 
\nonumber\\
&+& S_\mu S_\nu \phi^2 ) \biggr] - \frac{1}{2} \tilde g^{\mu\nu} \phi^2 \hat E_{\mu\nu},
\label{Choral-Lag}  
\end{eqnarray}
where we have defined $\hat E_{\mu\nu}$ as
\begin{eqnarray}
\hat E_{\mu\nu} = \frac{1}{2} \epsilon \partial_\mu \sigma \partial_\nu \sigma + \partial_\mu b_\nu 
+ i \partial_\mu \bar c_\lambda  \partial_\nu c^\lambda - \partial_\mu B \partial_\nu \sigma 
+ i \partial_\mu \bar c \partial_\nu c
+ ( \mu \leftrightarrow \nu ),
\label{hat-E}  
\end{eqnarray}
and used the relation (\ref{Dilaton}) between the scalar field $\phi$ and the dilaton $\sigma$.

Next, let us focus our attention on the last term in Eq. (\ref{Choral-Lag}) and rewrite it
into a more compact form:
\begin{eqnarray}
{\cal L}_q^{(E)} &\equiv& - \frac{1}{2} \tilde g^{\mu\nu} \phi^2 \hat E_{\mu\nu}
= - \frac{1}{2} \tilde g^{\mu\nu} \phi^2 \eta_{NM} \partial_\mu X^M \partial_\nu X^N
\nonumber\\
&=& - \frac{1}{2} \tilde g^{\mu\nu} \phi^2 \partial_\mu X^M \tilde \eta_{MN} \partial_\nu X^N.
\label{E-Lag}  
\end{eqnarray}
Here we have introduced an $IOSp(10|10)$ metric $\eta_{NM} = \eta_{MN}^T \equiv \tilde \eta_{MN}$ 
defined as \cite{Kugo}
\begin{align}
\eta_{NM} = \tilde \eta_{MN} =
\begin{array}{c}
x^\nu \\ b_\nu \\ \sigma \\ B \\ c^\nu \\ \bar c_\nu \\ c \\ \bar c
\end{array} &
\left(
\begin{array}{cc|cc|cc|cc}
     &                \delta_\mu^\nu &     &   &     &  \\ 
\delta^\mu_\nu  &                    &    &    &    &   \\ 
\hline
    &        &               \epsilon    &   -1      &     &    &      \\ 
    &        &               -1    &  0       &       &    &    \\
\hline   
    &        &     &    &       &   -i\delta_\mu^\nu  &   & \\  
    &        &     &    &   i\delta^\mu_\nu &   &    & \\
\hline
    &        &                &        &       &     &         &  -i \\  
    &        &                &        &       &     &          i     & \\
\end{array}
\right)_.
\label{OSp-metric}  \\
& \quad
\begin{array}{cccccccc}
x^\mu & b_\mu & \;\;\sigma & \; B & \;\;\, c^\mu & \;\;\, \bar c_\mu & \;\, c & \;\: \bar c
\end{array} \nonumber
\end{align}
Let us note that this $IOSp(10|10)$ metric $\eta_{NM}$, which is a c-number quantity, has the symmetry 
property such that 
\begin{eqnarray}
\eta_{MN}=(-)^{|M| \cdot |N|} \eta_{NM} = (-)^{|M|} \eta_{NM}=(-)^{|N|} \eta_{NM},
\label{Prop-OSp-metric}  
\end{eqnarray}
where the statistics index $|M|$ is 0 or 1 when $X^M$ is Grassmann-even or 
Grassmann-odd, respectively. This property comes from the fact that $\eta_{MN}$ is `diagonal' 
in the sense that its off-diagonal, Grassmann-even and Grassmann-odd, and vice versa, matrix elements 
vanish, i.e., $\eta_{MN} = 0$ when $|M| \neq |N|$, thereby being $|M| = |N| = |M| \cdot| N|$ in front of 
$\eta_{MN}$ \cite{Kugo}. 

Now that (\ref{E-Lag}) is expressed in a manifestly $IOSp(10|10)$ invariant form except for the Weyl invariant 
metric $\tilde g^{\mu\nu} \phi^2$, which will be discussed later, there could exist an $IOSp(10|10)$ as a global 
symmetry in our theory. Note that the infinitesimal $OSp$ rotation is defined by
\begin{eqnarray}
\delta X^M = \eta^{ML} \varepsilon_{LN} X^N \equiv \varepsilon^M{}_N X^N,
\label{OSp-rot}  
\end{eqnarray}
where $\eta^{MN}$ is the inverse matrix of $\eta_{MN}$, and the infinitesimal parameter
$\varepsilon_{MN}$ has the following properties:
\begin{eqnarray}
\varepsilon_{MN} = (-)^{1 + |M| \cdot |N|} \varepsilon_{NM}, \qquad
\varepsilon_{MN} X^L = (-)^{|L| (|M| + |N|)} X^L \varepsilon_{MN}.
\label{varepsilon}  
\end{eqnarray}
In order to find the conserved current, we assume that the infinitesimal parameter
$\varepsilon_{MN}$ depends on the space-time coordinates $x^\mu$, i.e., 
$\varepsilon_{MN} = \varepsilon_{MN} (x^\mu)$.

Assuming for a while that the metric $\tilde g^{\mu\nu} \phi^2$ is invariant under the $OSp$ 
rotation (\ref{OSp-rot}), we find that (\ref{E-Lag}) is transformed as
\begin{eqnarray}
\delta {\cal L}_q^{(E)} = - \tilde g^{\mu\nu} \phi^2 \left( \partial_\mu \varepsilon_{NM} 
X^M \partial_\nu X^N + \varepsilon_{NM} \partial_\mu X^M \partial_\nu X^N \right).
\label{Var-E-Lag}  
\end{eqnarray}
It is easy to prove that the second term on the RHS vanishes owing to the first property in 
Eq. (\ref{varepsilon}). Thus, ${\cal L}_q^{(E)}$ is invariant under the infinitesimal $OSp$ rotation.
The conserved current is then calculated to be:
\begin{eqnarray}
\delta {\cal L}_q^{(E)} &=& - \tilde g^{\mu\nu} \phi^2 \partial_\mu \varepsilon_{NM} X^M \partial_\nu X^N
\nonumber\\
&=& - \frac{1}{2} \tilde g^{\mu\nu} \phi^2 \partial_\mu \varepsilon_{NM} \left[ X^M \partial_\nu X^N
- (-)^{|M| \cdot |N|} X^N \partial_\nu X^M \right]
\nonumber\\
&=& - \frac{1}{2} \tilde g^{\mu\nu} \phi^2 \partial_\mu \varepsilon_{NM} \left( X^M \partial_\nu X^N
- \partial_\nu X^M X^N  \right)
\nonumber\\
&=& - \frac{1}{2} \tilde g^{\mu\nu} \phi^2 \partial_\mu \varepsilon_{NM} 
X^M \overset{\leftrightarrow}{\partial}_\nu X^N
\nonumber\\
&\equiv& - \frac{1}{2} \partial_\mu \varepsilon_{NM} {\cal M}^{\mu MN},
\label{OSp-current}  
\end{eqnarray}
with the conserved current ${\cal M}^{\mu MN}$ for the $OSp$ rotation taking the form: 
\begin{eqnarray}
{\cal M}^{\mu MN} = \tilde g^{\mu\nu} \phi^2 X^M \overset{\leftrightarrow}{\partial}_\nu X^N.
\label{OSp-current-M}  
\end{eqnarray}

The above proof makes sense only under the assumption that the metric $\tilde g^{\mu\nu} \phi^2$ 
and the other terms except for the last term in (\ref{Choral-Lag}) are invariant under the $OSp$ rotation, 
but it is obviously not the case. However, this problem is cured by noticing that the $OSp$ rotation 
includes a Weyl transformation on the dilaton:
\begin{eqnarray}
\delta \sigma = \eta^{\sigma L} \varepsilon_{LN} X^N = - \varepsilon_{BN} X^N \equiv - \varepsilon(x),
\label{Dilaton-OSp}  
\end{eqnarray}
where we have used (\ref{OSp-metric}) and
\begin{eqnarray} 
\begin{pmatrix}
   \epsilon & -1 \\
   -1 & 0
\end{pmatrix}^{-1}
= \begin{pmatrix}
   0 & -1 \\
   -1 & -\epsilon
\end{pmatrix},
\label{Matrix}  
\end{eqnarray}
where recall that the matrix $\eta^{ML}$ is the inverse matrix of $\eta_{ML}$. As for the scalar field $\phi(x)$, 
this transformation for the dilaton can be interpreted as a Weyl transformation:
\begin{eqnarray} 
\phi \rightarrow \phi^\prime = e^{- \varepsilon (x)} \phi. 
\label{Weyl-phi}  
\end{eqnarray}

Thus, simultaneously with the $OSp$ rotation, if we perform a Weyl transformation given by 
\begin{eqnarray} 
\delta g_{\mu\nu} = 2 \varepsilon (x) g_{\mu\nu},  \qquad
\delta S_\mu = - \partial_\mu \varepsilon (x),
\label{Weyl-g&S}  
\end{eqnarray}
and a local shift for the Nakanishi-Lautrup field $B$:\footnote{Under the $OSp$ rotation,
the $B$ field is transformed as $\delta B = \eta^{B L} \varepsilon_{LN} X^N = - \varepsilon_{\sigma N} X^N
+ \epsilon \varepsilon$. The transformation (\ref{B-shift}) is carried out independently of this
$OSp$ rotation.}   
\begin{eqnarray} 
\delta B = \epsilon \, \varepsilon (x),
\label{B-shift}  
\end{eqnarray}
it turns out that under the (local) $OSp$ rotation (\ref{OSp-rot}), the quantum Lagrangian ${\cal{L}}_q$
is transformed as
\begin{eqnarray}
\delta {\cal L}_q = - \frac{1}{2} \partial_\mu \varepsilon_{NM} {\cal M}^{\mu MN}.
\label{Var-q-Lag}  
\end{eqnarray}
As a result, the conserved current ${\cal M}^{\mu MN}$ for the $OSp$ rotation takes the form
(\ref{OSp-current-M}).

In a similar way, we can derive the conserved current for the infinitesimal translation:
\begin{eqnarray}
\delta X^M = \varepsilon^M,
\label{transl}  
\end{eqnarray}
and it turns out that the conserved current ${\cal P}^{\mu M}$ for the translation reads: 
\begin{eqnarray}
{\cal P}^{\mu M} = \tilde g^{\mu\nu} \phi^2 \partial_\nu X^M
= \tilde g^{\mu\nu} \phi^2 \left( 1 \overset{\leftrightarrow}{\partial}_\nu X^M \right).
\label{transl-current-P}  
\end{eqnarray}

From the conserved currents (\ref{OSp-current-M}) and (\ref{transl-current-P}), the corresponding conserved 
charges are given by
\begin{eqnarray}
M^{MN} &\equiv& \int d^3 x \, {\cal M}^{0 MN} = \int d^3 x \, \tilde g^{0 \nu} \phi^2  
X^M \overset{\leftrightarrow}{\partial}_\nu X^N,
\nonumber\\
P^M &\equiv& \int d^3 x \, {\cal P}^{0 M} = \int d^3 x \, \tilde g^{0 \nu} \phi^2 \partial_\nu X^M.
\label{IOSp-charge}  
\end{eqnarray}
For instance, the BRST charges for the GCT and Weyl transformation are respectively expressed as
\begin{eqnarray}
&{}& Q_B \equiv M (b_\rho, c^\rho) = \int d^3 x \, \tilde g^{0 \nu} \phi^2 
b_\rho \overset{\leftrightarrow}{\partial}_\nu c^\rho, 
\nonumber\\
&{}& \bar Q_B \equiv M (B, c) = \int d^3 x \, \tilde g^{0 \nu} \phi^2 
B \overset{\leftrightarrow}{\partial}_\nu c.
\label{Choral-Symm}  
\end{eqnarray}
We can then verify that using various ETCRs obtained so far, the $IOSp(10|10)$ generators $\{ M^{MN}, P^M \}$ 
generate an $IOSp(10|10)$ algebra:
\begin{eqnarray}
&{}& [ P^M, P^N \} = 0, 
\nonumber\\
&{}& [ M^{MN}, P^R \} = i \bigl[ P^M \tilde \eta^{NR} - (-)^{|N| |R|} P^N \tilde \eta^{MR} \bigr],
\nonumber\\
&{}& [ M^{MN}, M^{RS} \} = i \bigl[ M^{MS} \tilde \eta^{NR} - (-)^{|N| |R|} M^{MR} \tilde \eta^{NS} 
- (-)^{|N| |R|} M^{NS} \tilde \eta^{MR} 
\nonumber\\
&{}& + (-)^{|M| |R| + |N| |S|} M^{NR} \tilde \eta^{MS} \bigr].
\label{IOSp-algebra}  
\end{eqnarray}

Finally, it is useful to compare our extended choral symmetry $IOSp(10|10)$ 
with the original choral symmetry $IOSp(8|8)$ in Einstein's general relativity \cite{N-O-text}.
In our case, the choral symmetry is extended in the sense that the GCT is replaced with  
a larger symmetry, which consists of both the GCT and the Weyl gauge transformation.
Accordingly the dilaton $\sigma$, the Nakanishi-Lautrup field $B$, ghost $c$ and anti-ghost 
$\bar c$ are joined in the algebra. The choral symmetry $IOSp(10|10)$ therefore includes the
dilaton, or equivalently, the scalar field, which exists in the classical Lagrangian and is 
closely related to a classical theory. In contrast, the original $IOSp(8|8)$ symmetry 
is purely a symmetry among quantum fields, which are the NL field and ghosts, so the
symmetry is limited to the sector related to the gauge-fixing procedure. From this viewpoint,
we expect that the extended $IOSp(10|10)$ choral symmetry might play an important
role in clarifying the dynamics peculiar to the classical theory.

\section{Gravitational conformal symmetry and spontaneous symmetry breakdown}

One of the most interesting features in the formalism at hand is that as an analog of the well-known
conformal symmetry in a flat Minkowski space-time, there is a gravitational conformal symmetry
which is a subgroup of the choral symmetry, and its spontaneous symmetry breakdown down to the Poincar\'e 
symmetry guarantees that the graviton and the dilaton are exactly massless Nambu-Goldstone particles \cite{Oda-W}. 
This feature is so important for future developments of quantum gravity that we would like to explain 
the gravitational conformal symmetry and its spontaneous symmetry breakdown in detail. 

In particular, as already shown in Section 6, there is a $\it{massive}$ Weyl gauge field in the spectrum, 
so at first sight it appears to be strange that there is a conformal symmetry in the present theory since 
it is usually thought that conformal or scale symmetry exists in the theories with only massless particles.
With regard to this, it is worthwhile to recall that the massless Weyl gauge field acquires the mass via
spontaneous symmetry breakdown (SSB) of Weyl gauge symmetry and the SSB is the breakdown
of symmetry at the level of not  field operators but the representation of field operators in the sense
that the symmetry cannot be realized by a unitary transformation in the state vector space. Thus,
it is not strange that there is a conformal symmetry in the present theory with the massive gauge
field if the mass is generated through the SSB. Moreover, this physical situation is also supported by
the Zumino theorem \cite{Zumino} to some degree since the theorem insists that theories invariant
under general coordinate transformation and Weyl transformation at the same time should
possess conformal symmetry in a flat Minkowski background at least classically.
        
As clarified in the previous paper \cite{Oda-W}, the extended de Donder gauge condition (\ref{Ext-de-Donder}) 
and the scalar gauge condition (\ref{Scalar-gauge}) have a residual symmetry which corresponds 
to the dilatation and the special conformal transformation in a flat Minkowski space-time. Indeed,
the quantum Lagrangian (\ref{ST-q-Lag}) is still invariant under the restricted Weyl transformation \cite{Oda-R}:
\begin{eqnarray}
\delta g_{\mu\nu} &=& 2 \Lambda g_{\mu\nu},  \qquad
\delta \phi = - \Lambda \phi,
\nonumber\\
\delta S_\mu &=& - \partial_\mu \Lambda, \qquad
\delta b_\mu = - \partial_\mu \Lambda B,
\label{R-Weyl}  
\end{eqnarray}
where the infinitesimal transformation parameter $\Lambda$ takes the form:
\begin{eqnarray}
\Lambda = \lambda - 2 k_\mu x^\mu,
\label{R-Weyl-Lam}  
\end{eqnarray}
with $\lambda$ and $k_\mu$ being infinitesimal constants corresponding to a global scale 
transformation and the special conformal transformation, respectively \cite{Oda-W}.  Note that $\Lambda$ 
obeys the equation $g^{\mu\nu} \partial_\mu \partial_\nu \Lambda = 0$, which is a characteristic
feature of the restricted Weyl transformation. The whole global symmetry in the theory under
consideration should be included in the extended $IOSp(10|10)$ choral symmetry. Actually,
we can construct the generators corresponding to the transformation parameters 
$\lambda$ and $k_\mu$ out of those of the choral symmetry as
\begin{eqnarray}
D_0 &\equiv& - P(B) = - \int d^3 x \, \tilde g^{0 \nu} \phi^2 \partial_\nu B,
\nonumber\\
K^\mu &\equiv& 2 M^\mu (x, B) = 2 \int d^3 x \, \tilde g^{0 \nu} \phi^2 x^\mu
\overset{\leftrightarrow}{\partial}_\nu B.
\label{Res-gen}  
\end{eqnarray}
It is easy to verify that these generators generate the symmetry (\ref{R-Weyl}) in terms of the 
ETCRs in (\ref{ETCRs}). 

Our theory is also invariant under the translation and the general linear transformation $GL(4)$.
Actually, we can make the translation generator $P_\mu$ and $GL(4)$ generator $G^\mu{}_\nu$
from the choral symmetry as
\begin{eqnarray}
P_\mu &\equiv& P_\mu (b) = \int d^3 x \, \tilde g^{0 \nu} \phi^2 \partial_\nu b_\mu,
\nonumber\\
G^\mu{}_\nu &\equiv& M^\mu{}_\nu (x, b) - i M^\mu{}_\nu (c^\tau, \bar c_\tau)
\nonumber\\
&=& \int d^3 x \, \tilde g^{0 \lambda} \phi^2 ( x^\mu \overset{\leftrightarrow}{\partial}_\lambda b_\nu
- i c^\mu \overset{\leftrightarrow}{\partial}_\lambda \bar c_\nu ).
\label{Trans-GL}  
\end{eqnarray}
For instance, based on the ETCRs in (\ref{ETCRs}), we can check that the $GL(4)$ generator 
$G^\mu{}_\nu$ correctly generates the $GL(4)$ transformation on the fields $\phi, S_\rho$
and $g_{\sigma\tau}$:
\begin{eqnarray}
&{}& [ i G^\mu{}_\nu, \phi ] = x^\mu \partial_\nu \phi,  \qquad
[ i G^\mu{}_\nu, S_\rho ] = x^\mu \partial_\nu S_\rho + \delta_\rho^\mu S_\nu,
\nonumber\\
&{}& [ i G^\mu{}_\nu, g_{\sigma\tau} ] = x^\mu \partial_\nu g_{\sigma\tau} + \delta_\sigma^\mu g_{\nu\tau}
+ \delta_\tau^\mu g_{\nu\sigma}.
\label{Ex-GL}  
\end{eqnarray}

Finally, we can build a generator corresponding to the dilatation in a flat Minkowski space-time,
which is closely related to the generator $D_0$ of the scale transformation in (\ref{Res-gen}).
With this in mind, let us consider a set of generators, $\{ P_\mu, G^\mu{}_\nu, K^\mu, D_0 \}$. From these generators
we wish to construct the generator $D$ for the dilatation. Let us recall that in conformal field theory 
in the four-dimensional Minkowski space-time, the dilatation generator obeys the following algebra 
for an local operator $O_i (x)$ of conformal dimension $\Delta_i$ \cite{Gross, Nakayama}:
\begin{eqnarray}
[ i D, O_i (x) ] = x^\mu \partial_\mu O_i (x) + \Delta_i O_i (x).
\label{D-com}  
\end{eqnarray}
The scalar field $\phi (x)$, for example, has conformal dimension $1$ and therefore satisfies the equation:
\begin{eqnarray}
[ i D, \phi (x) ] = x^\mu \partial_\mu \phi (x) + \phi (x).
\label{D-phi-com}  
\end{eqnarray}
With this knowledge, let us construct a generator for the dilatation in such a way that the transformation law 
 on the scalar field satisfies this equation (\ref{D-phi-com}).
From Eq. (\ref{Ex-GL}) and the definition of $D_0$ in (\ref{Res-gen}), we find that 
\begin{eqnarray}
[ i G^\mu{}_\mu, \phi (x) ] = x^\mu \partial_\mu \phi (x), \qquad
[ i D_0, \phi (x) ] = - \phi (x).
\label{GD-phi-com}  
\end{eqnarray}
It therefore turns out that the following linear combination of $G^\mu{}_\mu$ and $D_0$ does the job:
\begin{eqnarray}
D \equiv G^\mu{}_\mu - D_0.
\label{D-def}  
\end{eqnarray}
As a consistency check, it is useful to see how this operator $D$ acts on the metric field.
The resulting expression is:
\begin{eqnarray}
&{}& [ i D, g_{\sigma\tau} ] = [ i G^\mu{}_\mu, g_{\sigma\tau} ] - [ i D_0, g_{\sigma\tau} ] 
\nonumber\\
&{}& = ( x^\mu \partial_\mu g_{\sigma\tau} + 2 g_{\sigma\tau} ) - 2 g_{\sigma\tau}
=  x^\mu \partial_\mu g_{\sigma\tau},
\label{D-g-com}  
\end{eqnarray}
which implies that the metric field has conformal dimension $0$ as defined in conformal field theory.
Further calculations reveal that the algebra among the generators $\{ P_\mu, G^\mu{}_\nu, K^\mu, D \}$
closes and takes the form:
\begin{eqnarray}
&{}& [ P_\mu, P_\nu ] = 0, \quad 
[ P_\mu, G^\rho{}_\sigma ] = i P_\sigma \delta^\rho_\mu, \quad
[ P_\mu, K^\nu ] = - 2 i ( G^\rho{}_\rho - D ) \delta^\nu_\mu, \quad
\nonumber\\
&{}& [ P_\mu, D ] = i P_\mu, \quad 
[ G^\mu{}_\nu, G^\rho{}_\sigma ] = i ( G^\mu{}_\sigma \delta^\rho_\nu - G^\rho{}_\nu \delta^\mu_\sigma),
\nonumber\\
&{}& [ G^\mu{}_\nu, K^\rho ] = i K^\mu \delta^\rho_\nu, \quad
[ G^\mu{}_\nu, D ] = [ K^\mu, K^\nu ] = 0, 
\nonumber\\
&{}& [ K^\mu, D ] = - i K^\mu, \quad
[ D, D] = 0. 
\label{Grav-conf0}  
\end{eqnarray}

To extract the gravitational conformal algebra in quantum gravity, it is necessary to introduce
the ``Lorentz'' generator. It can be contructed from the $GL(4)$ generator and the flat Minkowski metric 
to be:
\begin{eqnarray}
M_{\mu\nu} \equiv - \eta_{\mu\rho} G^\rho{}_\nu + \eta_{\nu\rho} G^\rho{}_\mu. 
\label{Lor-gene}  
\end{eqnarray}
In terms of the generator $M_{\mu\nu}$, the algebra (\ref{Grav-conf0}) can be cast to the form:
\begin{eqnarray}
&{}& [ P_\mu, P_\nu ] = 0, \quad 
[ P_\mu, M_{\rho\sigma} ] = i ( P_\rho \eta_{\mu\sigma} - P_\sigma \eta_{\mu\rho} ), 
\nonumber\\
&{}& [ P_\mu, K^\nu ] = - 2 i ( G^\rho{}_\rho - D ) \delta^\nu_\mu, \quad
[ P_\mu, D ] = i P_\mu, 
\nonumber\\
&{}& [ M_{\mu\nu}, M_{\rho\sigma} ] = - i ( M_{\mu\sigma} \eta_{\nu\rho} - M_{\nu\sigma} \eta_{\mu\rho}
+ M_{\rho\mu} \eta_{\sigma\nu} - M_{\rho\nu} \eta_{\sigma\mu}),
\nonumber\\
&{}& [ M_{\mu\nu}, K^\rho ] = i ( - K_\mu \delta^\rho_\nu + K_\nu \delta^\rho_\mu ), \quad
[ M_{\mu\nu}, D ] = [ K^\mu, K^\nu ] = 0, 
\nonumber\\
&{}& [ K^\mu, D ] = - i K^\mu, \quad
[ D, D] = 0. 
\label{Grav-conf}  
\end{eqnarray}
where we have defined $K_\mu \equiv \eta_{\mu\nu} K^\nu$. It is of interest that
the the algebra (\ref{Grav-conf}) in quantum gravity, which we call ``gravitational conformal algebra'', 
formally resembles conformal algebra in the flat Minkowski space-time except for the expression
of $[ P_\mu, K^\nu ]$ \cite{Gross, Nakayama}.\footnote{In case of conformal algebra in the flat space-time, 
the expression is given by $[ P_\mu, K^\nu ] = - 2 i ( \delta^\nu_\mu D + M_\mu{}^\nu )$.} This difference 
stems from the difference of the definition of conformal dimension (or weight) in both gravity 
and conformal field theory, for which the metric tensor field $g_{\mu\nu}$ has weight $2$ in gravity 
while it has weight $0$ in conformal field theory.
   
Now, on the basis of the gravitational conformal symmetry, we are able to show that $GL(4)$, 
special conformal symmetry and dilatation are spontaneously broken down to the Poincar\'e
symmetry. To this end, we postulate the existence of a unique vacuum $| 0 \rangle$, which is normalized 
to be the unity:
\begin{eqnarray}
\langle 0 | 0 \rangle = 1.
\label{Vac-norm}  
\end{eqnarray}
Furthermore, we assume that the vacuum is translation invariant:
\begin{eqnarray}
P_\mu | 0 \rangle = 0,
\label{Trans-Vac}  
\end{eqnarray}
and the vacuum expectation values (VEVs) of the metric tensor $g_{\mu\nu}$ and the scalar field $\phi$ are 
respectively the Minkowski metric $\eta_{\mu\nu}$ and a non-zero constant $\phi_0 \neq 0$:
\begin{eqnarray}
\langle 0 | g_{\mu\nu} | 0 \rangle = \eta_{\mu\nu}, \qquad
\langle 0 | \phi | 0 \rangle = \phi_0. 
\label{VEV-A}  
\end{eqnarray}
From (\ref{Ex-GL}),  we find that the VEV of an equal-time commutator between the $GL(4)$ generator 
and the metric field reads:
\begin{eqnarray}
\langle 0 | [ i G^\mu{}_\nu, g_{\sigma\tau} ] | 0 \rangle 
= \delta^\mu_\sigma \eta_{\nu\tau} + \delta^\mu_\tau \eta_{\nu\sigma}.
\label{G-g-CM}  
\end{eqnarray}
Thus, the Lorentz generator defined in Eq. (\ref{Lor-gene}) has a vanishing VEV:
\begin{eqnarray}
\langle 0 | [ i M_{\mu\nu}, g_{\sigma\tau} ] | 0 \rangle = 0.
\label{M-g-CM}  
\end{eqnarray}
On the other hand, the symmetric part defined as $\bar M_{\mu\nu} \equiv \eta_{\mu\rho}
G^\rho{}_\nu + \eta_{\nu\rho} G^\rho{}_\mu$ has the non-vanishing VEV: 
\begin{eqnarray}
\langle 0 | [ i \bar M_{\mu\nu}, g_{\sigma\tau} ] | 0 \rangle = 2 ( \eta_{\mu\sigma} \eta_{\nu\tau}
+ \eta_{\mu\tau} \eta_{\nu\sigma} ).
\label{BM-g-CM}  
\end{eqnarray}
Thus, the $GL(4)$ symmetry is spontaneously broken to the Lorentz symmetry where the corresponding 
Nambu-Goldstone boson with ten independent components is nothing but the massless graviton \cite{NO}.
Here, it is interesting that in a sector of the scalar field, the $GL(4)$ symmetry and of course the Lorentz
symmetry as well do not give rise to a symmetry breaking. This can be seen in the following commutators:
\begin{eqnarray}
\langle 0 | [ i G^\mu{}_\nu, \phi ] | 0 \rangle  = \langle 0 | [ i M_{\mu\nu}, \phi ] | 0 \rangle 
= \langle 0 | [ i \bar M_{\mu\nu}, \phi ] | 0 \rangle = 0.
\label{G-phi-CM}  
\end{eqnarray}
 
Now we wish to clarify how the dilatation and special conformal symmetry are spontaneously broken and
what the corresponding Nambu-Goldstone bosons are. As for the dilatation, we find that
\begin{eqnarray}
\langle 0 | [ i D, \sigma ] | 0 \rangle = 1,
\label{VEV-D-sigma}  
\end{eqnarray}
which elucidates the spontaneous symmetry breakdown of the dilatation whose
Nambu-Goldstone boson is just the massless dilaton $\sigma(x)$. 

Regarding the special conformal symmetry, we find:
\begin{eqnarray}
\langle 0 | [ i K^\mu, \partial_\nu \sigma ] | 0 \rangle = 2 \delta^\mu_\nu.
\label{VEV-K-phi}  
\end{eqnarray}
This equation means that the special conformal symmetry is certainly broken spontaneously
and its Nambu-Goldstone boson is the derivative of the dilaton. This interpretation can be
also verified from the gravitational conformal algebra as follows: In the algebra (\ref{Grav-conf}) we have
a commutator between $P_\mu$ and $K^\nu$:
\begin{eqnarray}
[ P_\mu, K^\nu ] = - 2 i ( G^\rho{}_\rho - D ) \delta^\nu_\mu.
\label{P-K}  
\end{eqnarray}
Let us consider the Jacobi identity:
\begin{eqnarray}
[ [ P_\mu, K^\nu ], \sigma ] + [ [ K^\nu, \sigma ], P_\mu ] + [ [ \sigma, P_\mu ], K^\nu ] = 0.
\label{Jacobi}  
\end{eqnarray}
Using the translational invariance of the vacuum in Eq. (\ref{Trans-Vac}) and the equation:
\begin{eqnarray}
[ P_\mu, \sigma ] =  - i \partial_\mu \sigma,
\label{Jacobi2}  
\end{eqnarray}
and taking the VEV of the Jacobi identity (\ref{Jacobi}), we can obtain the VEV:
\begin{eqnarray}
\langle 0 | [ K^\nu, \partial_\mu \sigma ] | 0 \rangle &=& - 2 \delta^\nu_\mu
\langle 0 | [ G^\rho{}_\rho - D, \sigma ] | 0 \rangle
\nonumber\\
&=& - 2 i \delta^\nu_\mu,
\label{VEV-Jacobi}  
\end{eqnarray}
which coincides with Eq. (\ref{VEV-K-phi}) as promised. 

In summary, the $GL(4)$ symmetry is spontaneously broken to the Poincar\'e symmetry whose Nambu-Goldstone
boson is the graviton. The dilatation symmetry and the special conformal symmetry are also
spontaneously broken and the corresponding Nambu-Goldstone bosons are the dilaton
and the derivative of the dilaton, respectively. Interest here is that the Nambu-Goldstone
boson associated with the special conformal symmetry is not an independent field in 
quantum gravity as in conformal field theory \cite{Kobayashi}.

\section{Conclusion}

In this article, we have presented a BRST formalism of a Weyl conformal gravity in Weyl geometry.
The essential ingredient in our formalism is choosing suitable gauge conditions for the general 
coordinate invariance and the Weyl invariance. To implement two independent BRST transformations
$\delta_B, \bar \delta_B$ corresponding to the GCT and the Weyl transformation, respectively, i.e.,
$\{ \delta_B, \bar \delta_B \} = 0$, one has to select the gauge conditions in such a way that 
the gauge condition for the GCT must be invariant under the Weyl transformation and 
that for the Weyl transformation must be so under the GCT \cite{Oda-W}. 

In addition, both gauge conditions must give us a gauge invariant measure in place of 
the conventional measure $\sqrt{-g}$ and ensure the masslessness of the dilaton. Interestingly enough,
such the gauge conditions are almost uniquely determined by the extended de Donder gauge condition 
(\ref{Ext-de-Donder}) for the GCT and the scalar gauge condition (\ref{Scalar-gauge}) for the Weyl 
transformation. With the other gauge conditions, we cannot construct the conserved currents for 
the extended choral symmetry, and without the choral symmetry we cannot ensure the gravitational 
conformal algebra such that we cannot prove the masslessness of the graviton and the dilaton. 
It is usually said that the gauge conditions do not change the physical content of a theory,
but it is true that the existence of global symmetries seems to critically depend on the gauge choice as 
seen in the present study of Weyl conformal gravity.

As for the future works, we would like to present a BRST formalism of quadratic conformal gravity 
(\ref{L-QG}) since this theory is the unique theory which is invariant under Weyl gauge transformation without
matter fields.\footnote{Recently, spontaneous symmetry breakdown of conformal symmetry in quantum 
quadratic gravity in Riemann geometry has been investigated in \cite{Kubo}.} However, it is known that 
higher-derivative gravities such as quadratic gravity generally suffer from the existence of 
a massless or massive ghost which prevents a lower bound of energy at the classical level 
and violates the unitarity at the quantum level. Thus, we have to provide a recipe for nullifying such a ghost. 
Since our choral symmetry is a huge global symmetry including 
the gravitational conformal symmetry, it might give us a useful tool for attacking various important problems 
such as the ghost and renormalizability.  The work is currently in progress with partial affirmative results.

\section*{Acknowledgment}

This work is supported in part by the JSPS Kakenhi Grant No. 21K03539 and P. S. is supported by
the IMPRS-PTFS.

\appendix
\addcontentsline{toc}{section}{Appendix~\ref{app:scripts}: Training Scripts}
\section*{Appendix}
\label{app:scripts}
\renewcommand{\theequation}{A.\arabic{equation}}
\setcounter{equation}{0}

\section{A derivation of constraints}
\def\T{\text{T}}

In this appendix we present a derivation of the secondary constraint (\ref{Second2}) and
show that the tertiary constraint is vanishing as seen in Eq. (\ref{Tert}).

Since the primary constraint (\ref{Primary}) is given by $\pi_S^0 \approx 0$, the terms
involving $S_0$ in the Hamiltonian density ${\cal{H}}_T$ contribute in the
calculation of the secondary constraint.  Thus, the relevant part in ${\cal{H}}_T$ is given by\footnote{We use 
the symbol ``$\sim$'' to denote the relevant terms.}
\begin{eqnarray}
{\cal{H}}_T \sim \pi_S^\mu \dot S_\mu + \pi_B \dot B - {\cal{L}}_q, 
\label{S-Ham}  
\end{eqnarray}
where the relevant part in ${\cal{L}}_q$ reads:
\begin{eqnarray}
{\cal L}_q &=& - 3 \xi \tilde g^{\mu\nu} \phi^2 S_\mu S_\nu + 6 \xi \tilde g^{\mu\nu} \phi S_\mu 
\partial_\nu \phi - \frac{1}{4} \sqrt{-g} H_{\mu\nu} H^{\mu\nu}
\nonumber\\
&-& \frac{1}{2} \epsilon \tilde g^{\mu\nu} D_\mu \phi D_\nu \phi
+ \tilde g^{\mu\nu} \partial_\mu B \phi \partial_\nu \phi.
\label{S-Lag}  
\end{eqnarray}
Note that $S_0$ in $- \frac{1}{4} \sqrt{-g} H_{\mu\nu} H^{\mu\nu}$ can be expressed in terms of
$\pi_S^i$ so this term can be ignored. 

Using Eqs. (\ref{CCM}), (\ref{dot-phi}) and (\ref{dot-many}), the Hamiltonian density takes the form:
\begin{eqnarray}
{\cal{H}}_T \sim \pi_S^i \partial_i S_0 + 3 \xi \tilde g^{\mu\nu} \phi^2 S_\mu S_\nu 
- 6 \xi \pi_B S_0 + \frac{1}{2} \epsilon \tilde g^{\mu\nu} D_\mu \phi D_\nu \phi,
\label{S-Ham2}  
\end{eqnarray}
where $\dot \phi$ is defined by (\ref{dot-phi}). Then, a straightforward calculation of (\ref{Second1})
gives us the secondary constraint (\ref{Second2}). 

Next, let us evaluate the tertiary constraint which comes from the time development of the
secondary constraint (\ref{Second2}). For this purpose, let us calculate the Poisson bracket
between the total Hamiltonian and each term in the secondary constraint (\ref{Second2}).
The method of the calculation is similar to that of the derivation of the secondary constraint;
just write out the relevant terms in the Hamiltonian density and then evaluate the Poisson bracket.
The results are presented in what follows:
\begin{eqnarray}
\{ H_T, \partial_i \pi_S^i \}_P &=& - ( 6 \xi + \epsilon ) \partial_i ( \tilde g^{i \mu} \phi D_\mu \phi ),
\nonumber\\
\{ H_T, \pi_B \}_P &=& \partial_i ( \tilde g^{i \mu} \phi \partial_\mu \phi ),
\nonumber\\
\{ H_T, \tilde g^{0 \mu} \phi^2 S_\mu \}_P &=& \partial_i ( \tilde g^{i \mu} \phi^2 S_\mu ).
\label{T-Ham1}  
\end{eqnarray}
From these expressions, it is easy to see that the tertiary constraint identically vanishes.
As a remark, a direct calculation of the last Poisson bracket is a bit complicated,
but instead we can make use of the relation between the Hamiltonian and time derivative
and the extended de Donder gauge condition (\ref{Ext-de-Donder}) and Eq. (\ref{S-eq}) as follows:
\begin{eqnarray}
\{ H_T, \tilde g^{0 \mu} \phi^2 S_\mu \}_P &=& - \partial_0 ( \tilde g^{0 \mu} \phi^2 S_\mu )
\nonumber\\
&=& - \partial_0 ( \tilde g^{0 \mu} \phi^2 ) S_\mu - \tilde g^{0 \mu} \phi^2 \partial_0 S_\mu
\nonumber\\
&=& \partial_i ( \tilde g^{i \mu} \phi^2 ) S_\mu + \tilde g^{i \mu} \phi^2 \partial_i S_\mu
\nonumber\\
&=& \partial_i ( \tilde g^{i \mu} \phi^2 S_\mu ).
\label{T-Ham2}  
\end{eqnarray}



\begin{thebibliography}{99}

\bibitem{Weyl}
H. Weyl, {``Gravitation und Elekrizit\"{a}t'', Sitzungsberichte der K\"{o}niglich Preu{\ss}ischen 
Akademie der Wissenschaften zu Berlin, 1918, pp. 465-480.}

\bibitem{Scholz}
E. Scholz, {``The Unexpected Resurgence of Weyl Geometry in late 20th-Century Physics'', 
Einstein Stud. {\bf 14} (2018) 261.}

\bibitem{Penrose}
R. Penrose, {``The Road to Reality'', Vintage Books, New York, 2007.}

\bibitem{Dirac}
P. A. M. Dirac, {``Long Range Forces and Broken Symmetries'', Proc. Roy. Soc. Lond. {\bf A 333} 
(1973) 403.}

\bibitem{Shirafuji}
K. Hayashi, M. Kasuya and T. Shirafuji, {``Elementary Particles and Weyl's Gauge Field'', Prog. Theor. Phys. 
{\bf 57} (1977) 431; Erratum: Prog. Theor. Phys. {\bf 59} (1978) 681.}

\bibitem{Hayashi}
K. Hayashi and T. Kugo, {``Remarks on Weyl's Gauge Field'', Prog. Theor. Phys. 
{\bf 61} (1979) 334.}

\bibitem{Cheng}
H. Cheng, {``Possible Existence of Weyl's Vector Meson'', 
Phys. Rev. Lett. {\bf 61} (1988) 2182.}

\bibitem{Percacci1}
C. Pagani and R. Percacci, {``Quantization and fixed points of non-integrable Weyl theory'', Class. Quant.
Grav. {\bf 31} (2014) 115005.}

\bibitem{Drechsler}
W. Drechsler and H. Tann, {``Broken Weyl Invariance and the Origin of Mass'', 
Found. Phys. {\bf 29} (1999) 1023.}

\bibitem{Ohanian}
H. C. Ohanian, {``Weyl Gauge-vector and Complex Dilaton Scalar for Conformal Symmetry 
and Its Breaking'', Gen. Rel. Grav. {\bf 48} (2016) 25.}

\bibitem{Cesare}
M. de Cesare, J. W. Moffat and M. Sakellariadou, {``Local Conformal Symmetry in
Non-Riemannian Geometry and the Origin of Physical Scales'',
Eur. Phys. J. {\bf C 77} (2017) 605.}  

\bibitem{Ghilencea1}
D. M. Ghilencea and H. M. Lee, {``Weyl Gauge Symmetry and Its Spontaneous Breaking in 
the Standard Model and Inflation'', Phys. Rev. {\bf D 99} (2019) 115007.}

\bibitem{Ghilencea2}
D. M. Ghilencea, {``Spontaneous Breaking of Weyl Quadratic Gravity to Einstein Action and
Higgs Potential'', JHEP {\bf 1903} (2019) 049.}

\bibitem{Oda-P}
I. Oda, {``Planck and Electroweak Scales Emerging from Weyl Conformal Gravity'',
PoS CORFU2018 (2019) 057, arXiv:1903.09309 [hep-th].}   

\bibitem{Oda-Corf}
I. Oda, {``Planck Scale from Broken Local Conformal Invariance in Weyl geometry'',
PoS CORFU2019 (2020) 070, arXiv:2003.12256 [hep-th].}

\bibitem{Bardeen}
W. A. Bardeen, {``On Naturalness in the Standard Model'', FERMILAB-CONF-95-391-T}.

\bibitem{Oda-Q}
I. Oda, {``Quantum Scale Invariant Gravity in de Donder Gauge'', Phys. Rev. {\bf D 105} (2022) 066001.}   

\bibitem{Oda-W}
I. Oda, {``Quantum Theory of Weyl Invariant Scalar-tensor Gravity'', Phys. Rev. {\bf D 105} (2022) 120618.}   

\bibitem{Oda-V}
I. Oda, {``Vanishing Noether Current in Weyl Invariant Gravity'', arXiv:2205.12517 [hep-th].}   

\bibitem{Nakanishi}
N. Nakanishi, {``Indefinite Metric Quantum Field Theory of General Gravity'', 
Prog. Theor. Phys. {\bf 59} (1978) 972.}

\bibitem{N-O-text}
N. Nakanishi and I. Ojima, {``Covariant Operator Formalism of Gauge Theories and Quantum Gravity'', 
World Scientific Publishing, 1990 and references therein.}

\bibitem{NO}
N. Nakanishi and I. Ojima, {``Proof of the Exact Masslessness of Gravitons'', 
Phys. Rev. Lett. {\bf 43} (1979) 91.}

\bibitem{MTW}
C. W. Misner, K. S. Thorne and J. A. Wheeler, {``Gravitation'', W H Freeman and Co (Sd), 1973.}

\bibitem{Kugo-Ojima}
T. Kugo and I. Ojima, {``Local Covariant Operator Formalism of Nonabelian Gauge Theories
and Quark Confinement Problem'', Prog. Theor. Phys. Suppl. {\bf 66} (1979) 1.}

\bibitem{Kugo}
T. Kugo, {``Noether Currents and Maxwell-type Equations of Motion in Higher Derivative Gravity Theories'', 
arXiv:2107.11600 [hep-th].}
        
\bibitem{Zumino}
B. Zumino, {``Effective Lagrangian and Broken Symmetries'', Lectures on Elementary Particles and 
Quantum Field Theory v.2, Cambridge, Brandeis Univ., pp. 437-500, 1970.}

\bibitem{Oda-R}
I. Oda, {``Restricted Weyl Symmetry'', Phys. Rev. {\bf D 102} (2020) 045008.}

\bibitem{Gross}
D. J. Gross and J. Wess, {``Scale Invariance, Conformal Invariance, and the High-Energy Behavior of
Scattering Amplitudes'', Phys. Rev. {\bf D 2} (1970) 753, and references therein.}

\bibitem{Nakayama}
Y. Nakayama, {``Scale Invariance Vs Conformal Invariance'', 
Phys. Rept. {\bf 569} (2015) 1, and references therein.}

\bibitem{Kobayashi}
K. Kobayashi and T. Uematsu, {``Non-linear Realization of Superconformal Symmetry'' 
Nucl. Phys. {\bf B 263} (1986) 309.}  

\bibitem{Kubo}
J. Kubo and J. Kuntz, {``Spontaneous Conformal Symmetry Breaking and Quantum Quadratic Gravity'', arXiv:2208.12832 [hep-th].}   






\end{thebibliography}
\end{document}